\documentclass[reprint,twocolumn,amsmath,amssymb,longbibliography]{revtex4-1}

\usepackage{graphicx}
\usepackage{subfigure}
\usepackage{longtable}
\usepackage{bm}
\usepackage{amsmath}
\usepackage{amssymb}
\usepackage{upgreek}
\usepackage{soul}  
\usepackage[rgb,dvipsnames]{xcolor}
\usepackage{physics}

\usepackage{siunitx} 

\usepackage{hyperref}	
\usepackage[capitalize]{cleveref} 

\DeclareSIUnit[]{\kelvin}{K}

\begin{document}

\title{Brillouin optomechanics in the quantum ground state}

\author{H. M. Doeleman$^*$}
\author{T. Schatteburg}
\thanks{These two authors contributed equally.}

\author{R. Benevides}
\author{S. Vollenweider}
\author{D. Macri}
\author{Y. Chu}
\email{yiwen.chu@phys.ethz.ch}
\affiliation{Department of Physics, ETH Z\"urich, Zurich, Switzerland}
\affiliation{Quantum Center, ETH Z\"urich, Zurich, Switzerland}

\begin{abstract} 
Bulk acoustic wave (BAW) resonators are attractive as intermediaries in a microwave-to-optical transducer, due to their long coherence times and controllable coupling to optical photons and superconducting qubits. However, for an optomechanical transducer to operate without detrimental added noise, the mechanical modes must be in the quantum ground state. 
This has proven challenging in recent demonstrations of transduction based on other types of mechanical resonators, where absorption of laser light caused heating of the phonon modes. 
In this work, we demonstrate ground state operation of a Brillouin optomechanical system composed of a quartz BAW resonator inside an optical cavity. The system is operated at $\sim$200 mK temperatures inside a dilution refrigerator, which is made possible by designing the system so that it self-aligns during cooldown and is relatively insensitive to mechanical vibrations. We show optomechanical coupling to several phonon modes and perform sideband asymmetry thermometry to demonstrate a thermal occupation below 0.5 phonons at base temperature. This constitutes the heaviest ($\sim\SI{494}{\micro\gram}$) mechanical object measured in the quantum ground state to date. Further measurements confirm a negligible effect of laser heating on this phonon occupation.
Our results pave the way toward low-noise, high-efficiency microwave-to-optical transduction based on BAW resonators. 
\end{abstract}

\maketitle

Quantum information processing has the potential to solve crucial problems that are beyond the reach of classical hardware \cite{Georgescu2014,Zoller2005}. Microwave-frequency solid state qubits, such as superconducting circuits \cite{Omalley2016,Wang2016,Jurcevic2021,Sivak2022} and spins in semiconductors \cite{Takeda2022,Philips2022}, have emerged as powerful platforms for quantum state manipulation, whereas optical photons are the natural choice for transporting quantum information over long distances \cite{Hensen2015,Pompili2021a}.
A microwave-to-optical conversion process \cite{Lauk2020,Lambert2020} would enable a modular, large- scale quantum processor \cite{Brecht2016} by connecting microwave qubits in physically separate dilution refrigerators through optical photons, which can be transported at room temperature through optical fibers. 
To be useful for quantum information processing, a microwave-to-optical transducer would have to be efficient and introduce fewer than 1 noise photon (referred to the input). 
One promising scheme uses a mechanical resonator as an intermediary to boost the effective electro-optical interaction \cite{Stannigel2010a,Chu2020}.
Over the past decade, such optomechanical transducers have been demonstrated with rapidly improving performance \cite{Bochmann2013,Higginbotham2018,Forsch2020,Jiang2020a,Stockill2022,Jiang2022,Weaver2022,Brubaker2022}, such as efficiencies up to 47\% using MHz frequency membranes as mechanical resonators \cite{Higginbotham2018,Brubaker2022} and added noise as low as 0.57 photons in devices based on GHz frequency phononic crystal resonators \cite{Mirhosseini2020}. 
Achieving high efficiency and low noise simultaneously, however, remains an outstanding challenge: MHz-frequency resonators suffer from thermal noise even at mK temperatures, whereas in phononic crystal resonators the circulating photon number, and therefore the efficiency, is limited by laser absorption and poor thermalization. 
Bulk acoustic wave (BAW) resonators have recently been shown to couple strongly to both superconducting qubits \cite{Chu2018a,VonLupke2022,Valimaa2022a} and infrared photons \cite{Renninger2018,Kharel2019,Kharel2022}, and form an attractive candidate for quantum transduction between the microwave and optical domains. They could potentially combine high efficiency transduction with low noise due to their high frequency and extremely low optical absorption \cite{Chu2020,Kharel2022}.
However, since previous optomechanical experiments on BAW resonators \cite{Renninger2018,Kharel2019,Kharel2022} were done at \SI{4}{\kelvin} with elevated phonon mode occupancy, it remains to be shown that BAW resonators can operate in the quantum ground state in the presence of the strong laser pump necessary to boost the optomechanical coupling rate.

Here, we report on ground-state operation of an optomechanically addressed BAW resonator.
The device is composed of of a quartz BAW resonator inside an optical Fabry-Perot cavity, similar to earlier systems measured at \SI{4}{\kelvin} \cite{Kharel2019,Kharel2022}.
Ground state operation at a thermal mode occupation of 0.3-0.4 phonons is achieved by operating at $\sim \SI{200}{\milli\kelvin}$ temperature inside a dilution refrigerator (hereafter abbreviated as fridge), which is made possible by  several modifications to improve alignment and stability against vibrations. 
Any efficient transduction process would require a pump laser power sufficient to reach an optomechanical cooperativity of unity. We show that our device remains in the ground state even under continuous illumination with pump powers that approach this regime, demonstrating its potential for simultaneously efficient and low-noise microwave-to-optical transduction. 
Finally, BAW resonators are also interesting for fundamental tests of physics due to their high mass and frequency. For example, a stricter bound on spontaneous wavefunction collapse rates could be obtained by measurement of an increasingly heavy mechanical resonator in as low a thermal occupation as possible \cite{Tobar2022}. To our knowledge, the measurement presented in this paper constitutes the heaviest mechanical resonator to date that was measured in the ground state.

\begin{figure}
    \centering
    \includegraphics{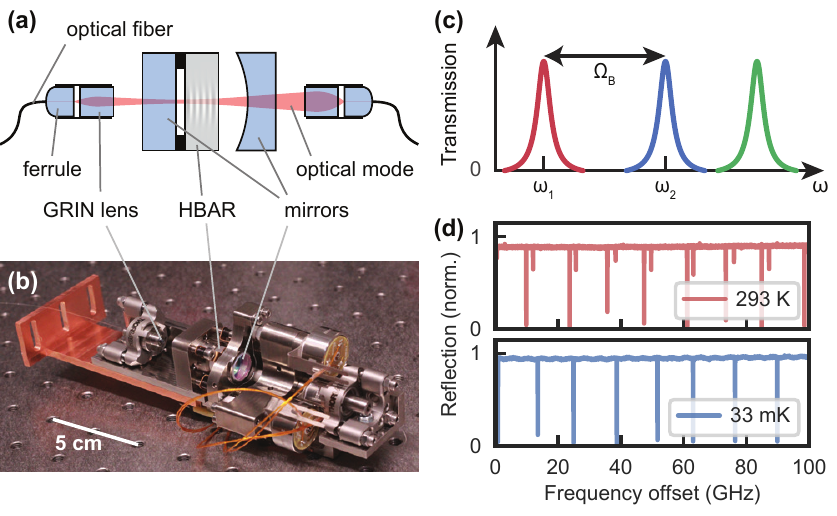}
    \caption{\textbf{Design and alignment of the cryogenic Brillouin cavity.}
    \textbf{(a)}~Schematic of the cryogenic Brillouin cavity.
    \textbf{(b)}~Picture of the assembled cavity. See text for detailed description.
    \textbf{(c)}~Frequency landscape. The optical modes are irregularly spaced due to reflections at the crystal surfaces. One mode pair is tuned to be separated by the Brillouin frequency $\Omega_\mathrm{B}$.
    \textbf{(d)}~Cavity reflection spectrum before (top) and after (bottom) cooldown, normalized to the maximum reflection when aligned by hand. The modes that approach zero reflection are the fundamental transverse modes. Higher-order Laguerre-Gaussian modes appear as shallower dips.
    }   \label{fig:1_CryoCavity}
\end{figure}

In our experiment, we use an optomechanical interaction between infrared photons in an optical cavity and acoustic waves in a BAW resonator.  The optical cavity has linewidth $\kappa/2\pi \approx \SI{2.4}{\mega\hertz}$
and is formed by a planar and a concave mirror, between which we place a $\SI{5}{\milli\meter}$ long z-cut quartz crystal with planar surfaces that acts as a high-overtone bulk acoustic wave resonator (HBAR, see \cref{fig:1_CryoCavity}a). The standing wave phonon modes are formed by reflections of the acoustic waves at the flat crystal interfaces to vacuum. This configuration leads to the Brillouin optomechanical coupling Hamiltonian
\begin{equation}
    \label{eq:OM_Hamiltonian}
    \hat{H}_\mathrm{int} = - \hbar g_{0,m} \qty(\hat{a}_1 \hat{a}_2^\dagger \hat{b}_m + \hat{a}_1^\dagger \hat{a}_2 \hat{b}_m^\dagger),
\end{equation}
with single-photon coupling rate $g_{0,m}$ between two optical modes $\hat{a}_{1/2}$ and a mechanical mode $\hat{b}_m$ \cite{Renninger2018,Kharel2019}. The interaction is caused by the interplay of electrostriction and photoelasticity of the crystal material. The electrostrictive effect allows the beat note between the two optical modes to create an elastic wave (the phonon mode). This elastic deformation in turn modifies the refractive index via the photoelastic effect, creating a grating that allows for Bragg scattering between the two optical modes. This interaction leads to the up-(down-)conversion of photons between the optical modes while simultaneously destroying (creating) a phonon. To enhance the interaction, a strong pump tone of classical intra-cavity amplitude $\alpha_p^\mathrm{cav}$ can be applied to the lower (higher) frequency optical mode, hereafter referred to as the red (blue) mode.
This effectively linearizes the Hamiltonian and leads to a beam-splitter (two-mode squeezing) type interaction with a cavity-enhanced coupling rate $g_m=g_{0,m}\alpha_p^\mathrm{cav}$. Since the interaction has to fulfill energy and momentum conservation, appreciable coupling is only observed for mechanical frequencies $\Omega_m$ near the Brillouin frequency, which for optical wavelengths of $\SI{\sim 1550}{nm}$ in quartz is $\Omega_\mathrm{B} /2\pi = \SI{12.65}{GHz}$.

 The implementation of this setup inside a fridge was accomplished by overcoming three outstanding challenges: coupling light in and out of the optical cavity in a dilution refrigerator, isolating the experiment from vibrations, and correcting for thermal misalignment. The optical modes of the cavity are addressed by light that is guided to and from the experiment by optical fibers. To both ferrules at the fiber ends, we glued a gradient-index (GRIN) lens at a specific distance that matches the outcoupled light to the cavity mode (see \cref{fig:1_CryoCavity}a). All optical components are placed on a compact steel mounting bracket that is material-matched to reduce thermal misalignment (see \cref{fig:1_CryoCavity}b). The whole bracket is then mounted to the base stage of a fridge via a system of springs for vibration isolation (see supplementary information, \cref{SI:Vibration_isolation}) and we place the cavity back mirror at a position that minimizes the effect of residual vibrations on the cavity mode frequency spacing (see supplementary information, \cref{SI:Displacement-insensitive_point}). We connect a dedicated thermometer to the steel mount holding the front mirror and the HBAR, which we will refer to later as the experiment thermometer. The in- and output lens mounts are at the ends of the supporting bracket and have to be operated by hand before cooldown. The mount holding the concave mirror of the cavity can be adjusted while the experiment is cold using stick-slip piezos. This is necessary to tune the frequency difference between the optical modes $\hat{a}_{1/2}$ to coincide with $\Omega_\mathrm{B}$ (see \cref{fig:1_CryoCavity}c), such that the optomechanical interaction is resonant. Note that the optical cavity modes are not spaced equidistantly due to reflections at the crystal interfaces \cite{Kharel2019}. Finally, to mitigate optical misalignment due to thermal contractions, we devised a method that lets the setup self-align during cooldown, the details of which are described in the supplementary information, \cref{SI:Alignment_procedure}. The results of the procedure are shown in \cref{fig:1_CryoCavity}d: At room temperature, the cavity reflection spectrum exhibits higher-order modes and a reduced reflection baseline compared to a reference measurement with optimum alignment, indicating slight misalignment. After the cooldown, the baseline increases to almost the value of the reference measurement. By additionally moving the back mirror mount to let the cavity mode overlap with the now aligned input beam, the higher-order modes disappear almost completely, showing that the cavity is completely aligned to the fundamental transverse mode. 


\begin{figure*}[t]
    \centering
    \includegraphics{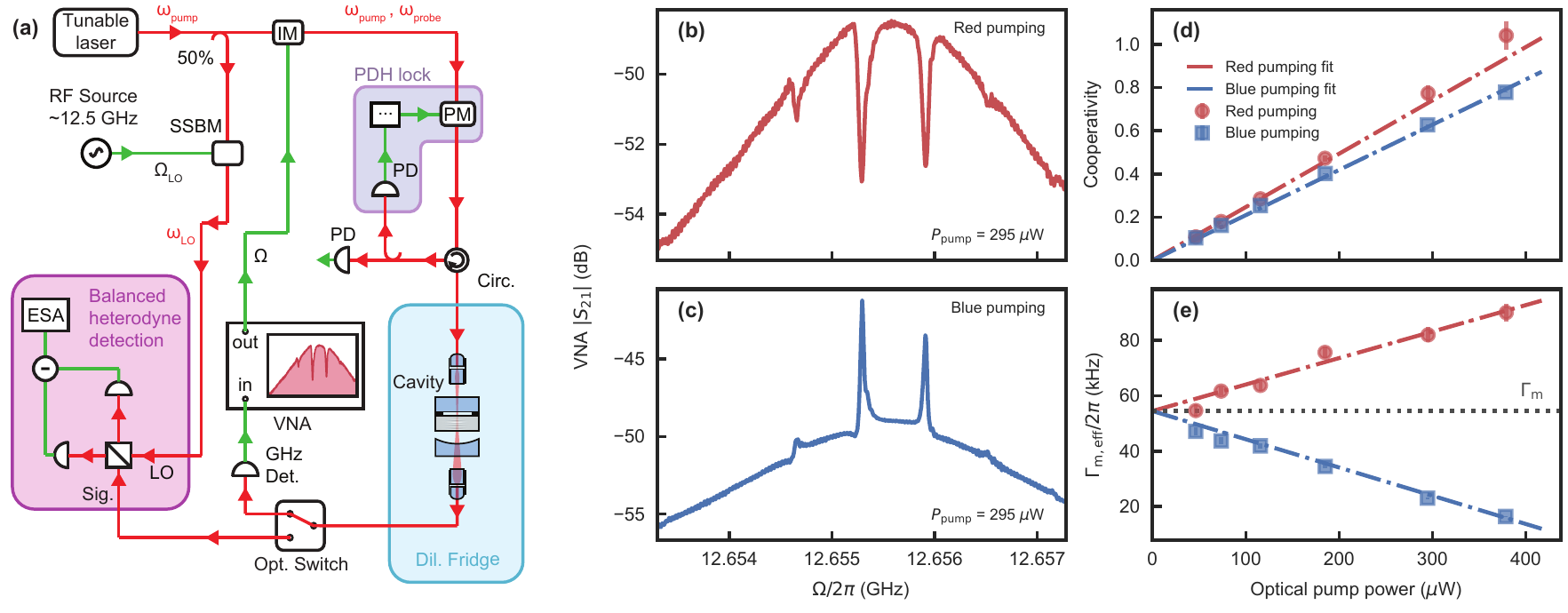}
    \caption{\textbf{Observing optomechanical interaction through OMIT and OMIA measurements.}
    \textbf{(a)}~Setup used for OMIT/OMIA measurements and sideband asymmetry thermometry. The output of a tunable diode laser is split into a signal and a LO arm and is locked on resonance with one of the optical modes using a Pound-Drever-Hall (PDH) lock. 
    When measuring OMIT/OMIA spectra, we use a vector network analyzer (VNA) to drive an intensity modulator (IM) in the signal arm, creating a weak probe tone detuned from the pump frequency $\omega_{\mathrm{pump}}$ by the VNA frequency $\Omega$. The beating of the transmitted pump and probe tones is recorded on a fast photodetector and sent to the VNA input. The transmission $S_{21}$ measured by the VNA is proportional to the probe transmission. 
    For sideband thermometry, no probe tone is created and the weak spontaneously scattered signal is measured through balanced heterodyne detection, using an electrical spectrum analyzer (ESA) and a strong LO that is frequency shifted by a single-sideband modulator (SSBM). An optical switch allows for rapid switching between these two configurations.
    \textbf{(b-c)}~Examples of OMIT (b) and OMIA (c) spectra measured at $\sim200$ mK. Three mechanical modes are clearly visible as narrow dips or peaks on top of a broad cavity transmission peak. 
    \textbf{(d-e)}~Linear scaling of optomechanical cooperativity $C_m$ (d) and effective mechanical linewidth $\Gamma_{\mathrm{m, eff}}$ (e) with pump power, for the center mechanical mode (mode 1) at $\sim$12.6553 GHz. Red and blue dash-dotted lines show linear fits. Black dotted line in (e) shows the intrinsic mechanical linewidth of $54.5\pm0.4$ kHz.
    }   \label{fig:2_OMITA}
\end{figure*} 

With the experimental setup compatible with the fridge environment, we characterize our BAW modes and their optomechanical coupling using optomechanically induced transparency (OMIT) and amplification (OMIA) measurements. In our triply-resonant scheme, an OMIT (OMIA) measurement is done by locking a strong pump laser to the red (blue) optical mode and sweeping a weak probe laser over the blue (red) mode~\cite{Kharel2019} (see \cref{fig:2_OMITA}a). The strong pump increases the cavity-enhanced coupling strength $g_m$, such that a narrow dip (peak) appears in the probe transmission spectrum when pump-probe detuning $\Omega$ equals a mechanical resonance frequency $\Omega_m$, as shown in \cref{fig:2_OMITA}b and c. 
The number of modes that show appreciable optomechanical coupling depends on the optical wavelength and cavity geometry~\cite{Kharel2019}. In this work we consider the three most prominent modes in our spectrum, labeled modes 0, 1 and 2 in increasing order of frequency. 
From fits to these spectra (see supplementary information, \cref{SI:OMITAfits}), we retrieve several system parameters, including the optomechanical cooperativity $C_m$ and the effective mechanical linewidth $\Gamma_{m, \mathrm{eff}}$. These show a linear dependence on pump laser power $P_p$ (see \cref{fig:2_OMITA}d,e), as expected from theory \cite{Kharel2019}. 
We extract intrinsic mechanical linewidths $\Gamma_m/2\pi$ of $\sim 50-55$ kHz for all modes, limited by diffraction loss, and estimate vacuum coupling rates $\abs{g_{0,m}}/2\pi$ of $\{4.02\pm 0.09 , 8.39 \pm 0.05 , 7.75 \pm 0.03 \}~\si{\hertz}$ for modes 0, 1 and 2, respectively. Both results are in good agreement with earlier measurements on flat-flat quartz crystals \cite{Kharel2019}. 

\begin{figure*}
    \centering
    \includegraphics{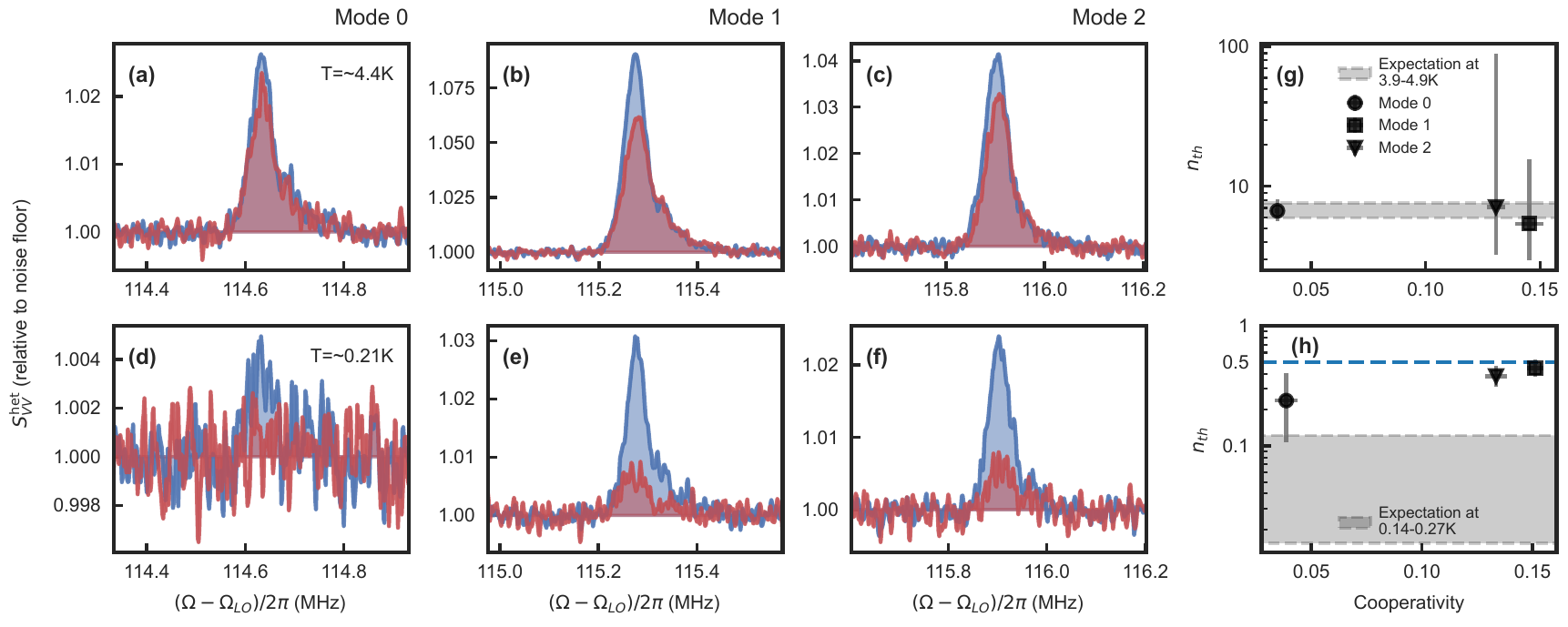}
    \caption{\textbf{Sideband asymmetry thermometry at $\sim \SI{4}{\kelvin}$ and $\sim \SI{200}{\milli\kelvin}$.}
    \textbf{(a-c)}~ESA spectra of the spontaneous Stokes (blue) and anti-Stokes (red) scattering at $T\sim \SI{4}{\kelvin}$ mediated by the three mechanical modes, measured with the pump laser locked to the blue and the red mode, respectively. Spectra are normalized to the noise floor and the red traces are corrected for several differences in pre-factors between red and blue pumped measurements (see supplementary information, \cref{SI:Thermometry_corrections}). Note that these frequencies are with respect to the LO, which is at a frequency $\Omega_{LO}$ roughly $\SI{115}{\mega\hertz}$ from the scattered signal frequencies. 
    \textbf{(d-f)}~Same as (a-c), but measured at $T \sim \SI{200}{\milli\kelvin}$. Here, the signal for mode 0 is weaker than that of modes 1 and 2 because of its lower coupling rate $g_0$ and because it is further off resonance from the optical mode (see \cref{fig:2_OMITA}b and c).
    \textbf{(g)}~Thermal mode occupations $n_{th}$ extracted from the red/blue asymmetries in (a-c) versus the mode cooperativity. The grey area indicates the expected occupation based on the experiment thermometer temperatures between start and end of the measurement.
    \textbf{(h)}~Same as (g), but for the measurements shown in (d-f). The blue dashed line indicates $n_{th}=0.5$.
    }   \label{fig:3_Thermometry}
\end{figure*}

Having established optomechanical coupling to several HBAR modes, we proceed to measure the thermal phonon occupations. This is done using optical sideband asymmetry thermometry, which relies on the difference that arises between Stokes and anti-Stokes scattering rates when the mechanical mode is near its quantum ground state \cite{Safavi-Naeini2012}. 
(Anti-)Stokes scattering corresponds to the second (first) term in \cref{eq:OM_Hamiltonian} and is associated with the creation (annihilation) of a phonon. Its probability scales proportionally with the thermal mode occupation as $n_{th} +1$ ($n_{th}$), reflecting the fact that it is impossible to destroy phonons if the resonator is in the ground state. Thus, by measuring the asymmetry between Stokes and anti-Stokes signals, $n_{th}$ can be determined. Note that $n_{th}$ refers to the occupation of the resonator in the absence of optomechanical backaction. 
While this technique has been frequently applied in various optomechanical systems \cite{Delic2019,Weinstein2014a,Peterson2016}, our system differs from these in that it uses not one but two optical modes: one that is resonant with the pump laser and the other with the Stokes or anti-Stokes signal. 
The expressions for these signals are therefore slightly modified and are presented in \cref{SI:Thermometry_theory} of the supplementary information. 
The mechanical sidebands are measured by locking the pump to one of the optical modes and mixing the cavity transmission with a frequency-shifted local oscillator (LO) in a balanced heterodyne detection setup (see \cref{fig:2_OMITA}a). Each measurement is performed in a $\sim \SI{4}{\min}$ time window during which the pulse tube of the cryostat is turned off to reduce vibrations.

We first verify the accuracy of our thermometry by measuring the mechanical modes at \SI{4}{\kelvin} without helium mix circulation. At this temperature, we expect the HBAR modes to be well thermalized to the surrounding environment because the \SI{4}{\kelvin} stage and all stages below it, as well as the experiment, are at the same temperature.
We perform measurements with the pump laser locked to the red or the blue mode, using sufficiently low pump power to ensure our optomechanical cooperativities are around 0.1, so as to minimize optomechanical backaction on the mechanical modes. 
Three thermal noise peaks appear at the mechanical frequencies of the three modes visible in \cref{fig:2_OMITA}b-c, shifted by the LO frequency (see \cref{fig:3_Thermometry}a-c). A clear asymmetry can be seen between the noise peaks in the cases of Stokes and anti-Stokes scattering.
Since we correct for other sources of asymmetry, such as cavity mode spacing drifts between the two measurements or the residual optomechanical backaction (see \cref{SI:Thermometry_corrections} of the supplementary information), the remaining asymmetry can be ascribed to the scaling with $n_{th} +1$ versus $n_{th}$ of the two scattering processes. The mode occupations we measure through the ratio of the areas under these peaks (see \cref{fig:3_Thermometry}g) show good agreement with the occupation of 5.9-7.6 phonons we expect based on the experiment thermometer readings throughout these measurements. 
To rule out the possibility that laser phase noise affects our thermometry, we measure the phase noise of our laser \cite{Kippenberg2013,Safavi-Naeini2013a} and find it to be sufficiently small to have a negligible influence on these results (see \cref{SI:PhaseNoise} of the supplementary information). 
Note that the uncertainty on the occupations of modes 1 and 2 are large because during the $\sim 4$-minute-long measurement, the cavity mode spacing shifts slightly due to heating, leading to an increased uncertainty on the detuning between the optical mode and mechanical modes. This affects modes 1 and 2 more than mode 0, as the former are situated on the flank of the optical resonance during the measurements at $\SI{4}{\kelvin}$ (see \cref{SI:Thermometry_OMIT_beforeafter} of the supplementary information for more details on our error sources). 
Furthermore, up until calculating the asymmetry between scattering signals, the errors are propagated via linear error propagation. However, because the function relating asymmetry and thermal mode occupation is strongly nonlinear in the range given by our error bars, the error bars for the thermal mode occupations shown in \cref{fig:3_Thermometry}g,h indicate the values corresponding to the extrema of the errorbars on the asymmetry.

We now cool our experiment down to $\sim \SI{200}{\milli\kelvin}$ and show that this brings the mechanical modes into the quantum ground state. At this lower temperature the noise peaks, shown in \cref{fig:3_Thermometry}d-f, decrease in amplitude and the asymmetry between red and blue pumping configurations increases.
The extracted occupations for modes 0,1 and 2 are
$0.24 \substack{+0.13 \\ -0.17}$, 
$0.44 \substack{+0.07 \\ -0.08}$ and 
$0.38 \substack{+0.07 \\ -0.08}$ phonons, respectively (see \cref{fig:3_Thermometry}h). 
In contrast to the \SI{4}{\kelvin} measurement, here the uncertainty is largest for mode 0 because it is at the flank of the optical resonance. 
All three modes are in the quantum ground state, with occupations $n_{th}<0.5$. 
Interestingly, these occupations are higher than what one would expect based on the thermometer readings, which predict occupations in the range of 0.015 to 0.12 phonons. 

\begin{figure}
    \centering
    \includegraphics{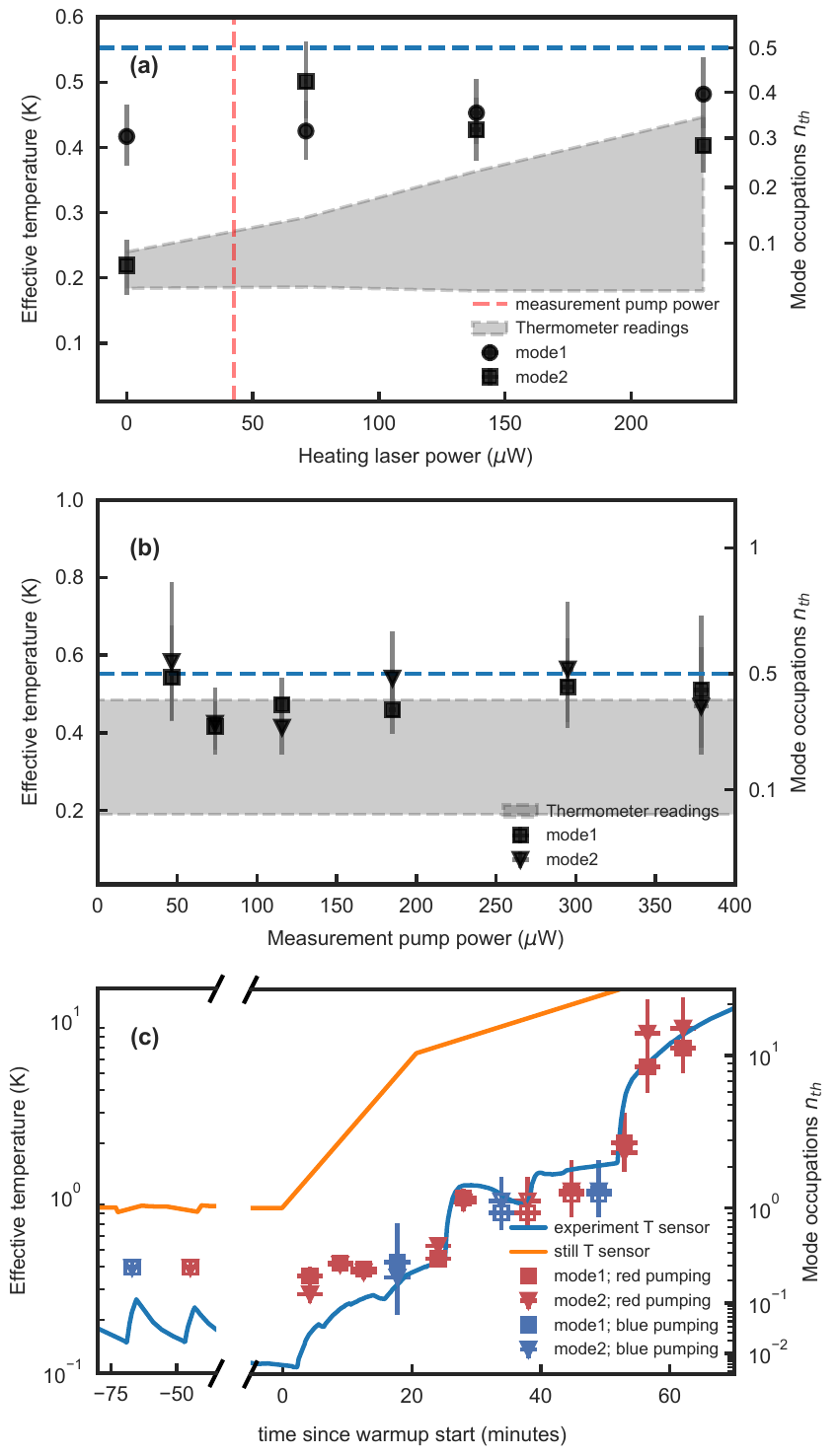}
    \caption{\textbf{Thermometry with laser or fridge heating.} Mode 0 is not shown due to its bad signal to noise ratio in these measurements.
    \textbf{(a)}~Thermometry with laser pre-heating. Effective mode temperatures after heating with the pump laser for 2 minutes, for different laser powers at the cavity input. The red dashed line indicates the pump power used during the thermometry measurements. The grey area indicates the measured temperature range between start and end of each measurement. The blue dashed lines in panels (a) and (b) indicate $n_{th}=0.5$.
    \textbf{(b)}~Pump power sweep. The measurement is similar to \cref{fig:3_Thermometry}h, but the \SI{4}{\minute} measurement window is split up into separate measurements with different pump powers. The grey area indicates the thermometer readings at the beginning and end of the entire sweep. Error bars are larger than in \cref{fig:3_Thermometry}h because integration time per measurement is shorter. 
    \textbf{(c)}~Thermometry during fridge warmup. Red (blue) markers show effective mode temperatures extracted from single red (blue) pumping measurements. The hollow markers indicate which measurements were used as reference pairs. Horizontal error bars indicate the measurement time of 4 minutes. Blue (orange) lines show the temperature sensor readings at the experiment (still flange). 
    }   \label{fig:4_LaserFridgeHeating}
\end{figure} 

A mode temperature above that of the crystal mount could arise from either laser absorption heating the crystal or heating from another source of radiation such as blackbody radiation from the higher-temperature stages.
To investigate these possible causes, we first repeat the sideband thermometry measurements at mK temperatures, but precede each measurement with two minutes where the pump laser is locked to the cavity and set to a power that is higher than during the thermometry measurement, effectively acting as a heat source. We find that the phonon mode occupations show no clear increase in thermal occupation with increasing heating laser power (see \cref{fig:4_LaserFridgeHeating}a), even though the crystal is subject to $\sim\SI{1}{W}$ of intra-cavity power at the highest heating power. The experiment thermometer temperature, however, increases significantly with laser power.
As a further test, and to rule out the possibility that the crystal cooled down in between the heating step and the thermometry measurement, we perform thermometry on our crystal using the same set of laser powers as shown in \cref{fig:2_OMITA}d-e. The resulting mode occupation shows no dependence on laser power (see \cref{fig:4_LaserFridgeHeating}b), demonstrating that laser absorption is not the cause of our elevated mode occupations.
The laser power responsible for the elevated experiment thermometer readings in \cref{fig:4_LaserFridgeHeating}a is therefore also not absorbed in the crystal. Instead, it is likely scattered due to imperfect alignment and eventually absorbed by other parts of the experiment. Since the experiment thermometer temperature never exceeds the effective phonon mode temperatures, however, this heating of the external environment does not have a significant effect on the mode temperatures. 

Having ruled out heating by laser absorption as source of elevated phonon mode temperatures, we then investigate the evolution of the mode temperature while the fridge warms up in order to test whether the heating is due to blackbody radiation from a higher temperature stage. To allow for faster measurements, we do this without relying on pairs of red and blue pumping measurements taken under the same conditions. Instead, we first extract the phonon occupations from a pair of reference measurements right before the warmup, and then use the fact that the corrected integrated signals from a subsequent red (blue) pumping measurement should scale as $n_{th}$ ($n_{th} + 1$) (see supplementary information, \cref{SI:Warmup_measurements}). Further reference pairs are taken during periods of stable temperature during the warmup, and used for subsequent measurements. We find that the temperatures agree well between phonon modes and follow the experiment temperature sensor, with the exception of low temperatures (see \cref{fig:4_LaserFridgeHeating}c). These observations are consistent with blackbody radiation from an additional heat source that starts at a higher temperature than the crystal mount before the warmup, but whose temperature increases more slowly than the still stage during the warmup. One possible culprit is the still shield, which surrounds the experiment and has a finite thermal conductivity to the still stage. In \cref{SI:ThermalModel} of the supplementary information, we discuss why blackbody radiation from the still stage or other stages directly does not explain our result.

We have demonstrated the operation of a cryogenic Brillouin cavity optomechanics system and used it to measure the modes of a BAW resonator in the quantum ground state. While the measurement of mechanical resonators in the quantum ground state has now been achieved in many mechanical systems, either through passive or active cooling \cite{Chu2020, Whittle2021}, the BAW modes studied here, with an effective mass of $\approx\SI{494}{\micro\gram}$, are to date the most massive mechanical objects measured with a thermal occupation of less than half a phonon (see supplementary information \cref{SI:EffMass} for a calculation of the effective mass). 
Our results represent an important step toward using BAW resonators for quantum transduction. We have overcome several crucial technical challenges, for example ensuring the alignment and stability of a free-space optomechanical cavity at mK temperatures. While the measured thermal occupations are higher than expected and not ideal for noiseless transduction, this is likely a particular issue of the current geometry. Importantly, we find no evidence of laser heating, and we point out that BAW resonators enclosed in microwave cavities have been measured to have much lower thermal occupations \cite{Chu2017}. Further improvements and upgrades to our system will include lower loss mechanical resonators, higher finesse optical cavities, and the incorporation of superconducting circuits. This work forms a solid foundation for these next steps toward a quantum transducer between the microwave and optical domains.

\section*{Acknowledgments}
The authors thank Matteo Fadel for calculating the effective mass of the mechanical modes, Maxwell Drimmer for his comments on the manuscript, Uwe von L\"upke for assisting in the initial phase of the project, Mona-Lisa Michel for work on the characterizing the vibration isolation stage, and Liu Qiu, Prash Kharel, Peter Rakich, David H\"alg, Thomas Gisler, Andrei Militaru and Johannes Piotrowski for helpful discussions.
We are grateful for support by the laboratory support group at the ETH physics department, as well as the mechanical workshop and design department for help with our cavity design and micromachining. 
H.M.D. was supported by an Rubicon grant (project no. 019.193EN.011) by the Dutch Research Council (NWO) for the duration of this project. This project has received funding from the European Research Council (ERC) under the European Union’s
Horizon 2020 research and innovation programme (grant agreement No 948047). 

\noindent {\bf Author contributions:}
 H.M.D. and T.S. built the experimental setup, performed measurements and analysis and developed the theory used in this work. R.B. helped perform and analyse laser phase noise measurements and wrote part of the corresponding section of the supplementary information. S.V. designed and built an early version of the optical cavity, built the 'dipstick' setup and helped develop the alignment procedure. D.M. designed the vibration isolation stage. Y.C. helped develop the theory and supervised the project. H.M.D., T.S. and Y.C. wrote the manuscript. 
 
\noindent {\bf Competing interests:}
The authors declare that they have no competing interests. 

\noindent {\bf Data and materials availability:} All data needed to evaluate the conclusions in the paper are present in the paper and/or the supplementary informations. Additional data available from authors upon request.

\bibliographystyle{apsrev4-1}
\bibliography{BOIQGS}

\onecolumngrid 
\newpage

\section*{Supplementary information}

\renewcommand{\thetable}{S\arabic{table}}  
\renewcommand{\thefigure}{S\arabic{figure}}
\renewcommand{\theequation}{S\arabic{equation}}
\setcounter{figure}{0}
\setcounter{table}{0}
\setcounter{section}{0}
\setcounter{equation}{0}

\subsection{Vibration isolation}
\label{SI:Vibration_isolation}

\begin{figure}[b]
    \centering
    \includegraphics{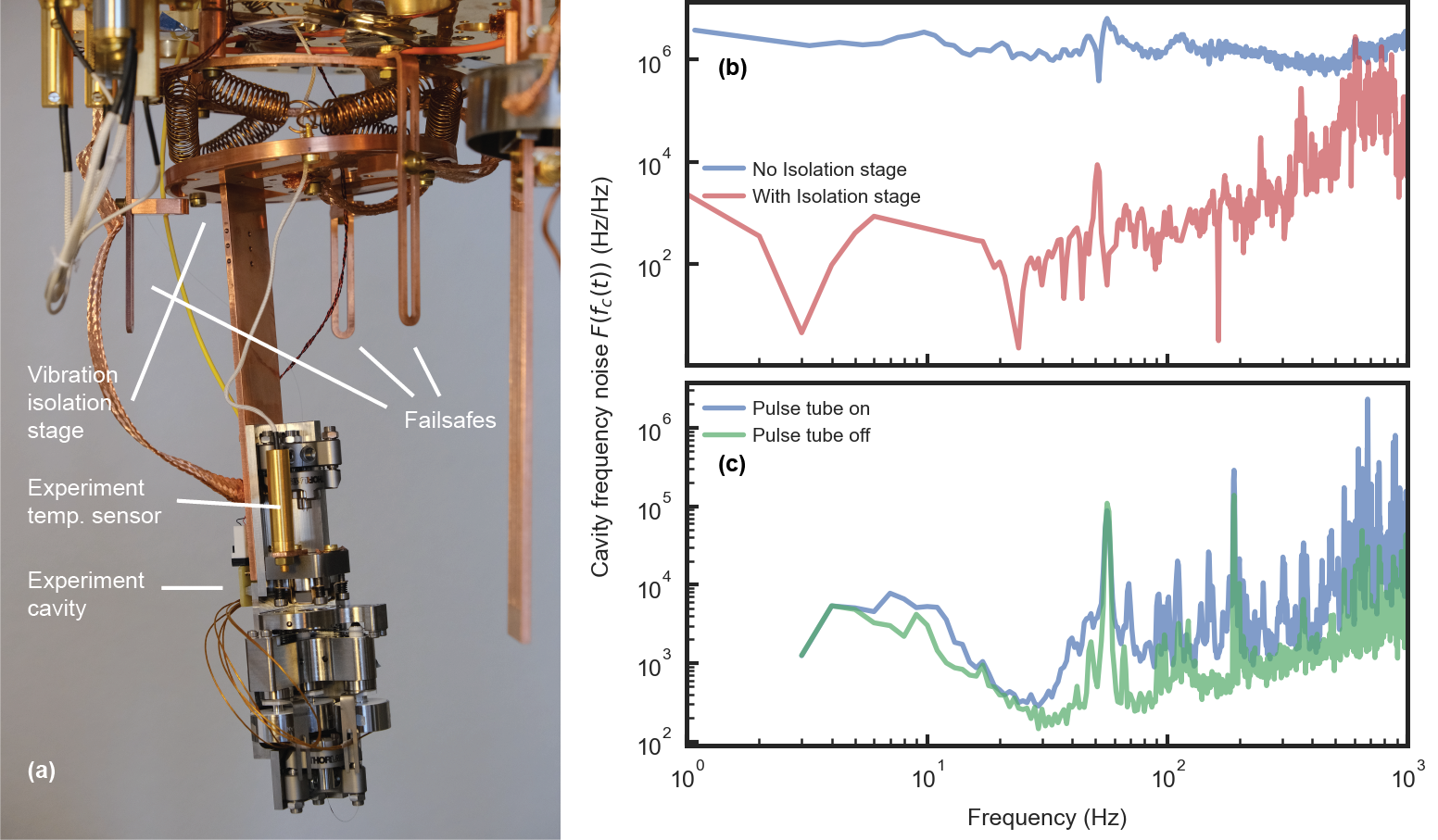}
    \caption{\textbf{Vibration isolation stage.}
    \textbf{(a)}~Experimental setup mounted in the fridge. The cavity is mounted on the vibration stage, which is suspended by springs from the mixing chamber plate of the fridge. Failsafes surround the vibration stage and serve to mount the experiment rigidly during manual alignment. The experiment temperature sensor is attached to the front mirror mount. 
    \textbf{(b)}~Vibration spectra recorded by the cavity at room temperature and with the fridge open, with and without vibration isolation. Vibrations were generated by a vibration speaker placed at the still plate. Background spectra, i.e. noise due to other fridge vibrations or ambient sound not caused by the speaker, were recorded and subtracted before plotting.  
    \textbf{(c)}~Vibration spectra recorded by the cavity in an evacuated fridge at \SI{4}{\kelvin} temperature, with and without the pulse tube running. 
    }   \label{SI:fig:Vibrations}
\end{figure} 

In contrast to an optical table, a dilution refrigerator is a mechanically very noisy environment. The main source of noise is the pulse tube, but there are also vibrations from the turbo pumps during mix circulation as well as other vibrations from the lab.
While this noise is mostly at frequencies below 1 kHz and therefore does not cause noise on our HBAR mechanical states, it does deteriorate our cavity lock quality and it causes noise on the optical mode spacing (see \cref{SI:Displacement-insensitive_point} for more details). 
We therefore mount our experiment onto a vibration isolation stage consisting of a plate suspended by six CuBe springs from the MXC stage (\cref{SI:fig:Vibrations}a). Loaded by the weight of the experiment, this mass-spring system has resonance frequencies between 1 and 10 Hz, and acts as a low-pass filter that suppresses mechanical vibrations above the resonance frequencies \cite{Lorenzelli2021Thesis}.

To test the effect of the isolation stage on the vibrations in our experiment, we perform time-resolved measurements of the cavity resonance frequency by sweeping a laser over one of our cavity resonance frequencies using a \SI{2}{\kilo\hertz} triangular sweep. The reflection spectrum is recorded on an oscilloscope and shows two resonance dips per sweep period. By finding the time differences between every second dip we obtain an array of time differences which would be equal to a sweep period if the cavity were perfectly stable, but in reality contain fluctuations due to the noise on the cavity frequency. We convert the time differences to frequencies using the known cavity linewidth as a frequency calibration and Fourier transform the array of frequencies to get the noise spectrum $F(f_c(t))(\omega)$ of the cavity frequency $f_c$ up to \SI{1}{\kilo\hertz}.
Note that a simple lock to the cavity resonance and recording of the error signal could give the same information, but locking was not possible in some of our measurements due to the large amount of noise.

The recorded noise spectra with and without isolation stage are shown in \cref{SI:fig:Vibrations}b, where the measurement without isolation is done by clamping the stage rigidly to the base plate using the failsafes shown in \cref{SI:fig:Vibrations}a. To generate reproducible vibrations, a sound containing all frequencies up to \SI{1}{\kilo\hertz} is played on a vibration speaker (Adin B1BT) placed on the still stage. These noise spectra show an attenuation of the noise by more than three orders of magnitude up to \SI{100}{\hertz}, whereas in the 300 to \SI{1000}{\hertz} range the attenuation is less. We speculate that there might be some higher order resonances or transmission through the thermal braids in that regime. 
Despite that, the total integrated frequency noise over the whole spectrum is reduced from \SI{1.5}{\giga\hertz} without stage to \SI{110}{\mega\hertz} with stage.

While the isolation stage is instrumental to reducing the vibration noise on our experiment, we still require the pulse tube to be turned off to be able to lock our laser to our cavity. With the pulse tube on, we measure a total integrated frequency noise of \SI{34}{\mega \hertz}, compared to just \SI{3.6}{\mega\hertz} with the pulse tube off (\cref{SI:fig:Vibrations}c). With the total noise commensurate with our cavity linewidth of $\sim$\SI{2}{\mega\hertz}, our cavity lock performs well. 
Finally, we note that during our measurements, we had to disconnect the control cables to the piezo motors on our back mirror from their driving modules because the voltage noise on the piezo control signals caused noticable noise on our cavity frequency. This is despite them being stick-slip piezos, which are kept at \SI{0}{\volt} when not moving.

\subsection{Displacement-insensitive point}
\label{SI:Displacement-insensitive_point}
Vibrations of the cavity mirrors not only affect the individual cavity resonance frequencies, but also the frequency spacing between these resonances. The former is mitigated by locking our laser to the pump mode, but this leaves the noise on the frequency spacing unaffected. All our measurements require that the frequency difference $\Delta_{12}$ between the two optical cavity modes is equal to the mechanical frequency $\Omega_m$ for the optomechanical interaction to be resonant. Thus, noise on the frequency spacing causes noise on the position of the broad optical resonance we observe in the OMIT and OMIA spectra. Since we average several spectra, this noise will result in a reduced height (depth) and a change of the lineshape for the OMIA (OMIT) features. Moreover, our thermometry signal strength at mechanical resonance $\Omega_m$ depends on $\Delta_{12}$ as a Lorentzian with the optical linewidth, peaking when $\Delta_{12}=\Omega_m$ (see \cref{SI:Thermometry_theory}). Noise on $\Delta_{12}$ therefore leads to a reduction of the averaged thermometry signals.

\begin{figure}[b]
    \centering
    \includegraphics{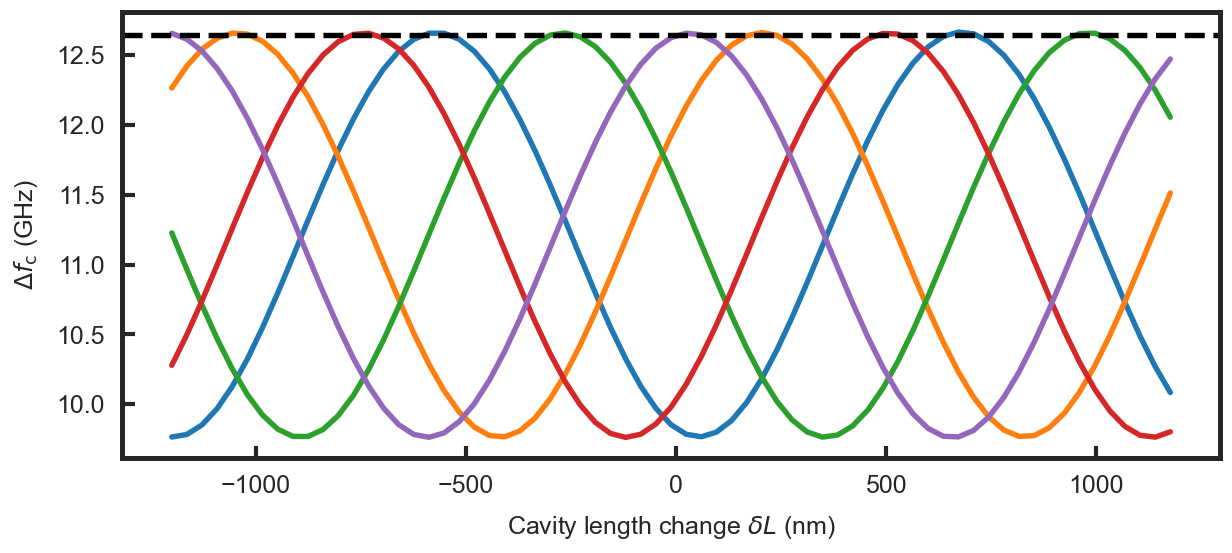}
    \caption{\textbf{Dependence of cavity mode spacing on cavity length.}
    The frequency spacings $\Delta f_\mathrm{c}$ between 6 neighbouring modes (shown as 5 lines with different colors) are shown as function of cavity length changes $\delta L$. Each mode pair shows an oscillatory dependence on $\delta L$ with a period of roughly \SI{1.3}{\micro\meter}. Dashed line indicates our Brillouin frequency.  We use cavity dimensions similar to those of our experiment. Since the exact value of the total physical cavity length is unknown in the experiment, we set it to \SI{10.4}{\milli\meter} to match the peak of the oscillations in $\Delta f_\mathrm{c}$ with the Brillouin frequency, as is the case in the experiment. This length corresponds to $\delta L=0$. To change $\delta L$, we change the distance between crystal and back mirror. 
    }   \label{SI:fig:DIP}
\end{figure}

In a vacuum-filled optical cavity of length $L$, the resonance frequency of the $m$-th mode is given by $f_{\mathrm{c},m}=\frac{mc}{2L}$, with $c$ the speed of light, while the mode spacing is given by $\Delta f_\mathrm{c}=\frac{c}{2L}$. In a $\sim$\SI{1}{\centi\meter} cavity at \SI{1550}{\nano\meter} wavelength, $m \sim 1.3\cdot10^4$. Thus, the dependence of $\Delta f_\mathrm{c}$ on small length changes $\delta L$ is $\sim 1.3\cdot10^4$ times weaker than that of the resonance frequency, and vibrations would have a neglegible effect. 
However, in our cavity, the reflections at the crystal interfaces lead to a strong dependence of mode spacing on both wavelength and mirror position, as observed in earlier work \cite{Kharel2019}. We use an analytical transmission matrix model, adapted from \cite{Kharel2019}, to calculate cavity reflection and transmission spectra, and from those find how $\Delta f_\mathrm{c}$ depends on small changes in cavity length (\cref{SI:fig:DIP}). This reveals that for our cavity geometry, $\Delta f_\mathrm{c}$ oscillates between $\sim$9.7 and \SI{12.7}{\giga\hertz}. The gradient $d(\Delta f_\mathrm{c})/d(\delta L)$ of this oscillation has a maximum value of \SI{7.5}{\mega\hertz/\nano\meter}, as compared with an average gradient $d(f_\mathrm{c})/d(\delta L)$ of the individual resonance frequencies of $\sim$\SI{15}{\mega\hertz/\nano\meter}. This shows that the mode spacing can depend nearly as strongly on mirror position in our system as the individual resonance frequencies.
We therefore tune our cavity length to the `displacement-insensitive point', where the Brillouin frequency coincides with the maximum of the mode spacing oscillation for one of the mode pairs in our spectrum. This corresponds to the situation shown in \cref{SI:fig:DIP}. At this point, the mode spacing is first order insensitive to mirror position, which greatly mitigates the noise on our measurements. We are able to match such a mode spacing maximum with the Brillouin frequency to within $\sim$\SI{1}{\mega\hertz} by adjusting the cavity length. 
 
\subsection{Alignment procedure}
\label{SI:Alignment_procedure}
\begin{figure}[b]
    \centering
    \includegraphics{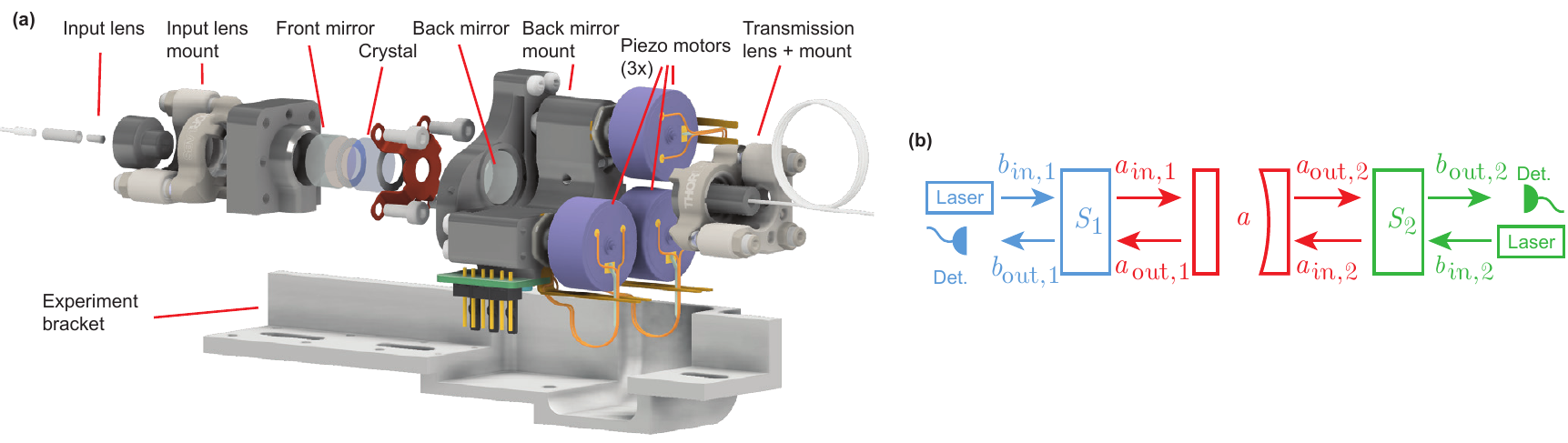}
    \caption{\textbf{Experiment design and description using scattering matrices.}
    \textbf{(a)}~Exploded view of the cavity design. All metal mounting parts are made of stainless steel to minimize relative movement during cooldown, except the back mirror mount which is made of titanium. The crystal is clamped to the front mirror using a CuBe clamp and using teflon spacers on each side.
    \textbf{(b)}~Schematic of our cavity and its input and output channels. Scattering matrix $S_1$ describes the coupling between the input optics and the cavity port 1, and $S_2$ that between the  transmission optics and the cavity port 2. 
    }   \label{SI:fig:Alignment_setup}
\end{figure}

During cooldown, the cryogenic components of our setup undergo thermal contraction, causing a misalignment between input and transmission optics and the cavity. While the experiment was designed to minimize such misalignments by matching materials and by fiber coupling rather than sending free-space beams through the fridge (see \cref{SI:fig:Alignment_setup}a), some misalignment remains. 
We therefore use a series of test cooldowns to determine the angular misalignments of the input and transmission lenses, and then we pre-compensate for these before cooling our experiment down in the dilution refrigerator. 
We will first discuss how cavity reflection and transmission spectra can be described in the case when input and output optics are not perfectly aligned to the cavity mode. We then present a model that parameterizes the effect of angular misalignment on the reflection and transmission spectra. This model is used to fit reflection and transmission spectra during the test cooldowns to find the room temperature settings for which the cavity is aligned at low temperatures (hereafter referred to as the 'cold optimum'). Next, we discuss how we test our model and fix some of its parameters by using room-temperature misalignments. Finally, we present the results of our test cooldowns and show that, if we position our cavity at the cold optimum, this leads to an improved alignment during cooldown.

\subsubsection{Misaligned cavity input-output theory}
\label{SI:Alignment:Theory1}

To understand the effect of misalignment on cavity transmission and reflection, we use cavity input-output theory, adapted to  describe the coupling between input or output optics with the cavity ports using scattering matrices (see \cref{SI:fig:Alignment_setup}b). This same theory will also be used in \cref{SI:Optical_input_coupling_characterization} to describe how we can determine input coupling rates from reflection spectra.
We consider a cavity and its two input-output ports (1 and 2), described as usual by input-output theory. 
The Langevin equation of motion in the harmonic basis for the classical amplitude $a$ of a cavity coupled to two such input-output channels is 
\begin{equation}
    -i \omega a =  -i \omega_0 a - \frac{\kappa}{2}a + \sqrt{\kappa^\mathrm{ext1}} a_\mathrm{in,1} + \sqrt{\kappa^\mathrm{ext2}} a_\mathrm{in,2}
\end{equation}
with $\omega_0$ the cavity resonance frequency and $\kappa=\kappa^\mathrm{ext1}+\kappa^\mathrm{ext2}+\kappa^\mathrm{int}$ is the loss rate, composed of the coupling rates $\kappa^\mathrm{ext1}$ and $\kappa^\mathrm{ext2}$ to ports 1 and 2 and intrinsic losses $\kappa^\mathrm{int}$. The input and output fields at port $i$ are written as $a_{\mathrm{in},i}$ and $a_{\mathrm{out},i}$, respectively.
The reflected field (at ports 1) and transmitted field (at port 2) are then given respectively as 
\begin{align}
    a_\mathrm{out,1} &= a_\mathrm{in,1} - \sqrt{\kappa^\mathrm{ext1}}a \\
    a_\mathrm{out,2} &= a_\mathrm{in,2} - \sqrt{\kappa^\mathrm{ext2}}a.
\end{align}

While $a_\mathrm{in}$ and $a_\mathrm{out}$ describe the input and output modes of the cavity, those do not, in general, correspond to the input and output modes of our input and transmission optics. We therefore introduce scattering matrices that couple these modes to each other in order to describe the mode mismatches due to misalignment, but also to include any further losses between the laser and cavity input, or cavity output and detector. Specifically, matrix $S_1$ couples the modes of our input optics $b_\mathrm{in,1}$ and $b_\mathrm{out,1}$ and the cavity modes $a_\mathrm{in,1}$ and $a_\mathrm{out,1}$. Similarly, $S_2$ couples the modes of our transmission optics $b_\mathrm{in,2}$ and $b_\mathrm{out,2}$ and the cavity modes $a_\mathrm{in,2}$ and $a_\mathrm{out,2}$. These are related by
\begin{align}
    \begin{pmatrix}
        b_\mathrm{out,1} \\
        a_\mathrm{in,1}
    \end{pmatrix}
    &= 
    \begin{pmatrix}
        s_{11,1} &  s_{12,1} \\
        s_{21,1} &  s_{22,1}
    \end{pmatrix}
    \begin{pmatrix}
        b_\mathrm{in,1} \\
        a_\mathrm{out,1}
    \end{pmatrix} \\
    \begin{pmatrix}
        b_\mathrm{out,2} \\
        a_\mathrm{in,2}
    \end{pmatrix}
    &= 
    \begin{pmatrix}
        s_{11,2} &  s_{12,2} \\
        s_{21,2} &  s_{22,2}
    \end{pmatrix}
    \begin{pmatrix}
        b_\mathrm{in,2} \\
        a_\mathrm{out,2}
    \end{pmatrix}
\end{align}
For simplicity, we assume that $s_{22,1}=s_{22,2}=0$, i.e. nothing coming out of the cavity gets reflected back into it (as this would create additional cavities and complicate analysis).

We now consider the situation where light is only inserted at port 1, such that $b_\mathrm{in,2}=a_\mathrm{in,2}=0$. Then we find that the reflection and transmission spectra are given by
\begin{align}
    R_1(\Delta) &= \left| \frac{b_\mathrm{out,1}}{b_\mathrm{in,1}} \right|^2 = 
    \left|s_{12,1} s_{21,1} \frac{\kappa/2 -i\Delta-\kappa^\mathrm{ext1}}{\kappa/2 -i\Delta} + s_{11,1} \right|^2 , \label{SI:eq:R1}\\
    T_1(\Delta) &= \left| \frac{b_\mathrm{out,2}}{b_\mathrm{in,1}} \right|^2 = 
    \left|s_{12,2} s_{21,1} \right|^2  \frac{\kappa^\mathrm{ext1} \kappa^\mathrm{ext2}}{\kappa^2/4 +\Delta^2}, \label{SI:eq:T1}
\end{align}
with $\Delta=\omega-\omega_0$. The reflection and transmission coefficients $R_2$ and $ T_2$ for the situation where light is inserted only at port 2 are obtained by swapping the final indices 1 and 2. 

\subsubsection{The effect of angular misalignments on cavity reflection and transmission}
\label{SI:Alignment:Theory2}
Having established a framework for how misalignments may affect the cavity reflection and transmission spectra, we now proceed to find closed-form expressions for reflection and transmission as function of input and transmission lens tilt angles. At the end of this section, we present the procedure that we use to determine the optimal input and transmission lens tilt angles at low temperature. 

The input lens is well aligned when the input beam reflects off the first (flat) mirror under normal incidence, such that the reflected light (off resonance from the cavity modes, i.e. for $|\Delta|\gg \kappa$) is directed back into the fiber. When this condition is met, the cavity mode can be spatially aligned in the x-y plane (mirror surface) to the position where the input beam hits the front mirror by tilting the curved back mirror, which displaces the cavity modes in x and y. 
An angular misalignment $d\theta_\mathrm{in}, d\phi_\mathrm{in}$ of the input lens leads to a displacement of the reflected beam from the center of the fiber. As both the fundamental mode of the fiber and the reflected beam are Gaussian of shape, and as the overlap integral of two Gaussians is a Gaussian as well, the reflection coefficient is expected to depend on these misalignments as 
\begin{equation}
    R_{1,|\Delta|\gg \kappa}(d\theta_\mathrm{in},d\phi_\mathrm{in}) = R_{1,\mathrm{max}} e^{-A^2(d\theta_\mathrm{in}^2+d\phi_\mathrm{in}^2)/2 \theta_0^2}, \label{SI:eq:R1_2}
\end{equation}
where $\theta_0$ is the angle for which the beam is displaced by roughly one cavity waist on the front mirror, $A$ is a dimensionless fit parameter that we can determine by room temperature misalignment tests and $R_{1,\mathrm{max}}$ is the reflected power at perfect alignment (and $|\Delta|\gg \kappa$). Note that $\theta_0$ is redundant, and we therefore fix it to  $50^\circ$, such that $A$ is of order 1. To write the overlap integral in this way, we assumed that the phase front mismatch is negligible. From \cref{SI:eq:R1} we see that $R_{1,|\Delta|\gg \kappa}=\left|s_{12,1} s_{21,1} + s_{11,1} \right|^2$, so input misalignments result in a change of $s_{12,1}, s_{21,1}$ and $s_{11,1}$.

The transmission lens is well aligned when the transmitted power measured at the end of the fiber is maximized. Ideally, this would involve both an optimization of the angle and the position of the transmission lens, but the limited space in our setup does not allow for independent control of both. Therefore, we only control the lens mount tilt, which affects both angle and position. 
From \cref{SI:eq:T1} we see that, in so far as they affect transmission, misalignments must be reflected by a change in $s_{21,1}$ and $s_{12,2}$, where $s_{21,1}$ captures misalignments of the input lens and $s_{12,2}$ those of the transmission optics.

Since $s_{21,1}$ describes how much of the input mode $b_\mathrm{in,1}$ is converted to the cavity input mode $a_\mathrm{in,1}$, we take it to be proportional to the overlap integral of the respective mode fields $E_{a,1}$ and $E_{b,1}$ on the front mirror surface.
Taking this surface to be the $z=0$ plane, and assuming the input beam and cavity mode to have the same waist size $w_0$, the fields $E_{a,1}$ and $E_{b,1}$ are respectively described by the fields of a Gaussian beam and a tilted and displaced Gaussian beam, i.e.
\begin{align}
    E_{a,1}(x,y,z) &= E_{a,1} \, e^{-(x^2+y^2)/w_0^2}e^{i k_z z}, \label{SI:eq:Ea}\\
    E_{b,1}(x,y,z) &= E_{b,1} e^{-((x-\Delta x -d\theta_\mathrm{in}z)^2+(y-\Delta y - d\phi_\mathrm{in}z)^2)/w_0^2} \, e^{i k_z (z+d\theta_\mathrm{in}(x+\Delta x)+d\phi_\mathrm{in}(y+\Delta y))}. \label{SI:eq:Eb}
\end{align}
Here, $d\theta_\mathrm{in}$ and $d\phi_\mathrm{in}$ are the input beam tilt angles in the $xz$- and $yz$-plane, respectively, with respect to the $z$-axis. We have assumed them to be small, such that $\sin(d\theta_\mathrm{in})\approx d\theta_\mathrm{in}$ and $\sin(d\phi_\mathrm{in})\approx d\phi_\mathrm{in}$. 
The displacements $\Delta x$ and $\Delta y$ are those between the input beam and the cavity mode on the front mirror, i.e. $\Delta x = \Delta x_\mathrm{in} - \Delta x_\mathrm{c}$ and $\Delta y = \Delta y_\mathrm{in} - \Delta y_\mathrm{c}$, 
with $\{\Delta x_\mathrm{in},\Delta y_\mathrm{in}\}$ and $\{\Delta x_\mathrm{c},\Delta y_\mathrm{c}\}$ the displacements of the input beam and cavity mode, respectively, on the front mirror with respect to the point of optimum alignment. 
In \cref{SI:eq:Ea,SI:eq:Eb}, we have also taken the beam waists to be at the mirror and ignored the Gaussian beam divergence (our input beam has a Rayleigh range of more than $\sim$\SI{10}{\milli\meter}). 
Due to the large distance $d_\mathrm{IL}\sim\SI{22}{\milli\meter}$ between the input lens and the front mirror and the fact that we consider displacements on the order of the cavity waist (\SI{77}{\micro\meter}), our tilts will be of order $\arctan{(77\cdot10^{-3}/22)}= \SI{3e-3}{\radian}$ and we may safely ignore the tilt-dependent terms in $E_{b,1}(x,y,z)$. The field $E_{b,1}(x,y,z)$ is thus simply a Gaussian diplaced by $\Delta x$ and $\Delta y$, which we can rewrite as a function of the tilts $\{d\theta_\mathrm{in}, d\phi_\mathrm{in} \}$ by defining 
\begin{align}
    \Delta x_\mathrm{in} &= B d_\mathrm{IL} d\theta_\mathrm{in}, \\
    \Delta y_\mathrm{in} &= B d_\mathrm{IL} d\phi_\mathrm{in}, \\
    \Delta x_\mathrm{c} &= C d_\mathrm{IL} d\theta_\mathrm{bm}, \\
    \Delta y_\mathrm{c} &= C d_\mathrm{IL} d\phi_\mathrm{bm}, \\
    w_0 &= d_\mathrm{IL} \theta_0, 
\end{align}
where $B$ and $C$, like $A$ before, are dimensionless fit parameters we can determine from room temperature tests, and $\{d\theta_\mathrm{bm}, d\phi_\mathrm{bm}\}$ are the tilt angles of the concave back mirror. 
Note that in reality $\Delta x_\mathrm{c}$ and $\Delta x_\mathrm{c}$ do not depend on $d_\mathrm{IL}$ but rather on the radius of curvature of the back mirror, and including that correction would simply lead to a different value for $C$. 
Using these definitions, and calculating the overlap integral of $E_{a,1}$ and $E_{b,1}$ in the $z=0$ plane, we then find that $s_{21,1}$, normalized to the value at perfect alignment ($d\theta_\mathrm{in}=d\phi_\mathrm{in}=d\theta_\mathrm{bm}=d\phi_\mathrm{bm}=0$) shows a simple Gaussian dependence on tilts, described by
\begin{align}
    \frac{s_{21,1}(d\theta_\mathrm{in}, d\phi_\mathrm{in},d\theta_\mathrm{bm}, d\phi_\mathrm{bm})}{s_{21,1}(0,0,0,0)} &= \frac{\iint_{x,y}E_{a,1}(x,y) E_{b,1}^*(x,y,d\theta_\mathrm{in}, d\phi_\mathrm{in},d\theta_\mathrm{bm}, d\phi_\mathrm{bm}) dxdy}{\iint_{x,y}E_{a,1}(x,y) E_{a,1}^*(x,y,0,0,0,0) dxdy} \notag \\
    &= e^{-(dx^2+dy^2)/2w_0^2} \notag \\
    &= e^{-\left((B d\theta_\mathrm{in}-C d\theta_\mathrm{bm})^2+(B d\phi_\mathrm{in}-C d\phi_\mathrm{bm})^2\right)/2\theta_0^2}. \label{SI:eq:s21-1}
\end{align}

With the same arguments as used to get to \cref{SI:eq:s21-1}, the dependence of the scattering matrix element $s_{12,2}$, which describes how much of the cavity output mode $a_\mathrm{out,2}$ is converted to the output mode $b_\mathrm{out,2}$ on the transmission side of the cavity, can be written as
\begin{equation}
    \frac{s_{12,2}(d\theta_\mathrm{tr}, d\phi_\mathrm{tr},d\theta_\mathrm{bm}, d\phi_\mathrm{bm})}{s_{21,1}(0,0,0,0)} =
    e^{-\left((D d\theta_\mathrm{tr}-E d\theta_\mathrm{bm})^2+(D d\phi_\mathrm{tr}-E d\phi_\mathrm{bm})^2\right)/2\theta_0^2}, \label{SI:eq:s12-2}
\end{equation}
with $\{d\theta_\mathrm{tr},d\phi_\mathrm{tr}\}$ the tilt angles of the transmission lens and $D$ and $E$ two more dimensionless fit parameters. 
Plugging \cref{SI:eq:s21-1,SI:eq:s12-2} into \cref{SI:eq:T1} and evaluating at cavity resonance ($\Delta=0$), we find
\begin{equation}
    T_{1,\Delta=0} 
    = T_{1,\mathrm{max}} e^{-\left((B d\theta_\mathrm{in}-C d\theta_\mathrm{bm})^2+(B d\phi_\mathrm{in}-C d\phi_\mathrm{bm})^2 
                + (D d\theta_\mathrm{tr}-E d\theta_\mathrm{bm})^2+(D d\phi_\mathrm{tr}-E d\phi_\mathrm{bm})^2 \right)/\theta_0^2}, \label{SI:eq:T1_2}
\end{equation}
where $T_{1,\mathrm{max}}$ is the resonant transmission at perfect alignment. 

All the tilt angles used in \cref{SI:eq:R1_2,SI:eq:s21-1,SI:eq:s12-2} are defined with respect to the point of optimal alignment, i.e.
\begin{align}
    d\theta_\mathrm{in} &= \theta_\mathrm{in}-\theta_\mathrm{in,0}, \\
    d\theta_\mathrm{bm} &= \theta_\mathrm{bm}-\theta_\mathrm{bm,0}, \\
    d\theta_\mathrm{tr} &= \theta_\mathrm{tr}-\theta_\mathrm{tr,0},    
\end{align}
and similarly for the y-tilts $d\phi_\mathrm{in},d\phi_\mathrm{bm}$ and $d\phi_\mathrm{tr}$. 
Note that the optimal tilts $\{\theta_\mathrm{in,0}, \phi_\mathrm{in,0}, \theta_\mathrm{bm,0}, \phi_\mathrm{bm,0}, \theta_\mathrm{tr,0}, \phi_\mathrm{tr,0}\}$ generally depend on temperature due to the thermal contractions. 

Because the back mirror piezo motors that control $\{\theta_\mathrm{bm},\phi_\mathrm{bm}\}$ are open loop and suffer from hysteresis and unequal step sizes in their two directions of movement, as well as a change of step size with temperature, we cannot generally know the back mirror tilts. We can only know when the back mirror is positioned such that the cavity mode is perfectly aligned to the input beam, i.e. $\Delta x_\mathrm{in}=\Delta y_\mathrm{in}=0$, by finding the position for which higher-order Laguerre-Gaussian modes disappear from our reflection spectrum. 
If we take both our input lens and the back mirror to be perfectly aligned, such that $d\theta_\mathrm{in}=d\phi_\mathrm{in}=d\theta_\mathrm{bm}=d\phi_\mathrm{bm}=0$, \cref{SI:eq:T1_2} simplifies to:
\begin{equation}
    T_{1,\Delta=0} (d\theta_\mathrm{tr}, d\phi_\mathrm{tr}) = T_{1,\mathrm{max}} e^{-D^2\left( d\theta_\mathrm{tr}^2 + d\phi_\mathrm{tr}^2 \right)/\theta_0^2}. \label{SI:eq:T1_3}
\end{equation}

\Cref{SI:eq:R1_2,SI:eq:T1_3} can be used to find the optimal input lens and transmission lens alignments at low temperatures, using the following procedure:
\begin{enumerate}
    \item At room temperature, determine the parameter $A$ in \cref{SI:eq:R1_2} by measuring reflection spectra for a set of controlled misalignments $\{ d\theta_\mathrm{in},d\phi_\mathrm{in} \}$ of the input lens, and fitting average off-resonant reflection values to \cref{SI:eq:R1_2}.
    \item At room temperature, determine the parameter $D$ in \cref{SI:eq:T1_3} by measuring transmission spectra for a set of controlled misalignments $\{ d\theta_\mathrm{tr},d\phi_\mathrm{tr} \}$ of the transmission lens (with input lens and back mirror perfectly aligned), and fitting average off-resonant transmission values to \cref{SI:eq:T1_3}.
    \item Set the input lens to its room-temperature optimum, i.e. where $d\theta_\mathrm{in}=d\phi_\mathrm{in}=0$, by maximizing off-resonant reflection. Cool down the cavity and measure reflection spectrum. Repeat such cooldowns for four other input lens alignment settings. 
    \item Fit the resulting five low-temperature off-resonant reflection values to \cref{SI:eq:R1_2}, to obtain the input lens cold optimum $\{\theta_\mathrm{in,0},\phi_\mathrm{in,0}\}^\mathrm{cold}$.
    \item With the input lens set to cold optimum $\{\theta_\mathrm{in,0},\phi_\mathrm{in,0}\}^\mathrm{cold}$, cool down and align back mirror in-situ. Record transmission spectrum. Repeat this step for five different transmission lens alignment settings.
    \item Fit the resulting five low-temperature resonant transmission values to \cref{SI:eq:T1_3}, to obtain the transmission lens cold optimum $\{\theta_\mathrm{tr,0},\phi_\mathrm{tr,0}\}^\mathrm{cold}$.
\end{enumerate}


\subsubsection{Room temperature calibrations}
\label{SI:Alignment:RTcal}
Here we discuss the results of the room-temperature calibrations done to determine parameters $A$ and $D$ in \cref{SI:eq:R1_2,SI:eq:T1_3}, i.e. steps 1 and 2 of the procedure listed at the end of \cref{SI:Alignment:Theory2}.
By constraining our fits, we need only a minimum of 3 cooldowns to determine the input cold optimum, and another 3 for the transmission lens cold optimum. This is preferred over fitting more parameters on a larger cooldown dataset, because the cooldowns are most time-consuming. 

We align our cavity manually to the warm optimum, i.e. $d\theta_\mathrm{in}=d\phi_\mathrm{in}=d\theta_\mathrm{bm}=d\phi_\mathrm{bm}=d\theta_\mathrm{tr}=d\phi_\mathrm{tr}=0$ at room temperature. 
To fit for $A$, we then record reflection spectra at 1550-1552 nm wavelength at a set of controlled misalignments $\{ d\theta_\mathrm{in},d\phi_\mathrm{in} \}$ of the input lens, centered at the optimum. Throughout this work, input and transmission lens tilts are measured in degrees rotation of the adjustment screws on the tip-tilt stages of these lenses, and a $10^\circ$ rotation corresponds to \SI{32}{\micro\radian} physical tilt of the lens. \cref{SI:fig:Alignment}a shows the resulting fit of \cref{SI:eq:R1_2} to the average off-resonant reflection values, normalized to reflection at perfect alignment, at all these misalignment positions. This reflection follows the expected Gaussian dependence.
To fit for $D$, we keep the input lens and back mirror at their optima and record transmission spectra at a set of controlled misalignments $\{ d\theta_\mathrm{tr},d\phi_\mathrm{tr} \}$ of the transmission lens, centered at the optimum. The fit of the average resonant transmission values, normalized to resonant transmission at perfect alignment,  to \cref{SI:eq:T1_3} also shows good agreement with a Gaussian (see \cref{SI:fig:Alignment}b). 

\subsubsection{Finding the cold optimum using cooldowns}
\label{SI:Alignment:cooldowns}

\begin{figure}
    \centering
    \includegraphics{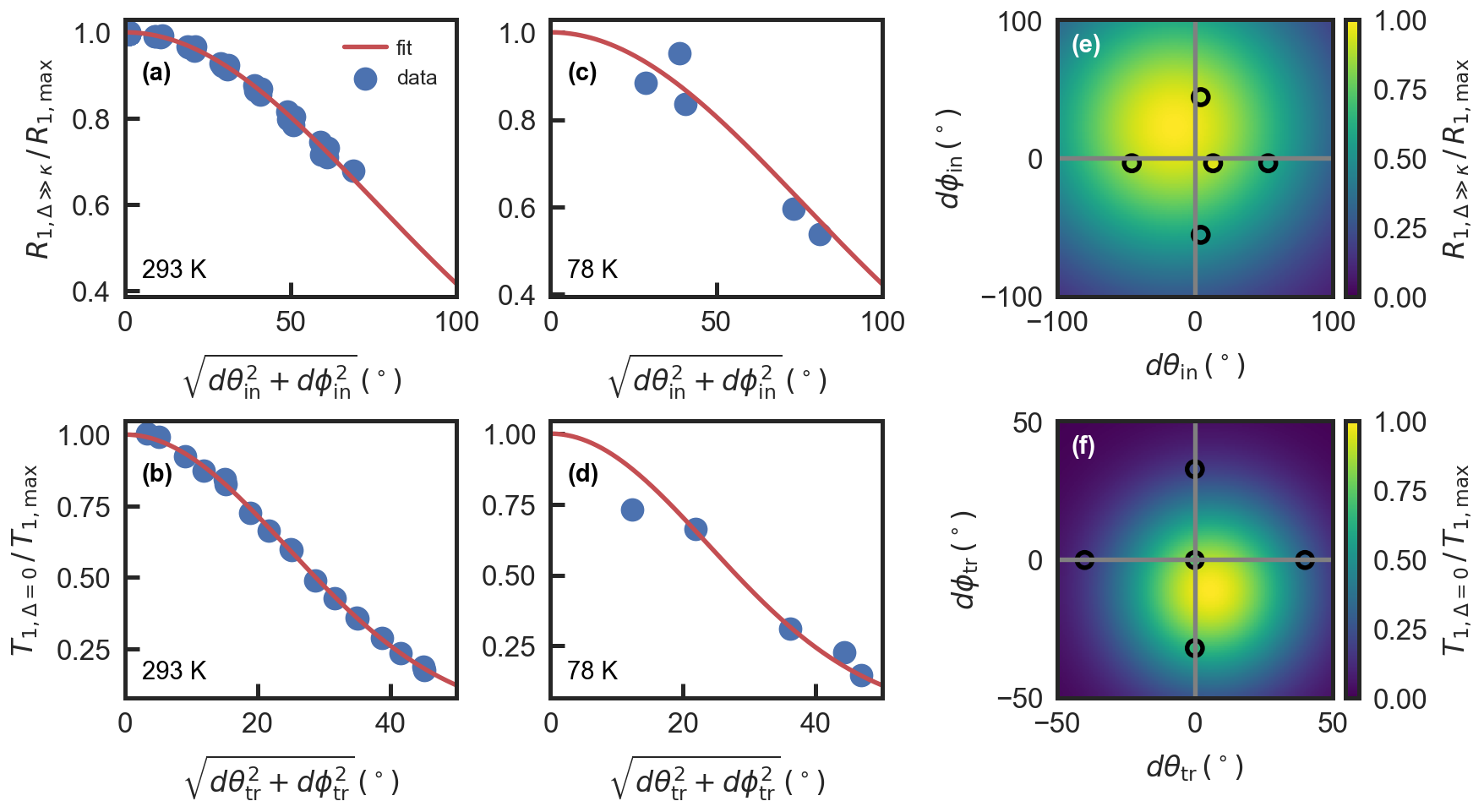}
    \caption{\textbf{Determination and compensation of thermal misalignments.}
    \textbf{(a)}~Calibration fit of off-resonant reflection to \cref{SI:eq:R1_2} for several room temperature input lens misalignments. From this fit we obtain fit parameter $A$, which sets the Gaussian width. 
    \textbf{(b)}~Calibration fit of resonant transmission to \cref{SI:eq:T1_3} for several room temperature transmission lens misalignments. From this fit we obtain $D$, which sets the Gaussian width. 
    \textbf{(c)}~Off-resonant reflection fit for data taken at \SI{78}{\kelvin} temperature, during 5 cooldowns with different input lens misalignments. This fit produces the input lens cold optimum.
    \textbf{(d)}~Resonant reflection fit for data taken at \SI{78}{\kelvin} temperature, during 5 cooldowns with different transmission lens misalignments (and with input lens and back mirror at cold optimum). This fit produces the transmission lens cold optimum.
    \textbf{(e)}~Same as (c), but plotted as function of $d\theta_\mathrm{in}$ and $d\phi_\mathrm{in}$ separately, and with the origin set to the warm optimum. Color map shows the fit, circles show the data. The warm-to-cold optimum shift is the difference between the color map maximum and the origin.
    \textbf{(f)}~Same as (e), but plotted as function of $d\theta_\mathrm{tr}$ and $d\phi_\mathrm{tr}$ separately, and with the origin set to the warm optimum. 
    }   \label{SI:fig:Alignment}
\end{figure} 

 Here we discuss the results of the cooldowns used to determine input lens and transmission lens cold optima, i.e. steps 3-6 of the procedure listed at the end of \cref{SI:Alignment:Theory2}. 
 The cooldowns are done in a liquid nitrogen dipstick, consisting of a $\sim\SI{1.5}{\meter}$ tube with a chamber at the bottom hosting our experiment and vacuum, electrical and fiber connections at the top. The bottom half of the dipstick is then immersed in liquid nitrogen.
A temperature sensor is mounted onto the cavity and after evacuation, helium is added to the chamber for faster thermalization. The experiment typically requires $\sim\SI{30}{\minute}$ to reach a temperature of \SI{78}{\kelvin}. This is still far above the base temperature of a dilution fridge, but we found that most thermal contraction happens in this first $\sim\SI{220}{\kelvin}$ drop, consistent with the strong decrease of the thermal expansion coefficient of stainless steel around this temperature\cite{Pobell2007}.
A reflection and transmission spectrum is recorded every minute during cooldown, but here we only use the spectra taken at \SI{78}{\kelvin}. One can, however, use the full dataset to track the optimal alignment during the cooldown. After reaching base temperature, the dipstick is pulled out of the nitrogen dewar, warmed up and vented once above \SI{0}{\degreeCelsius}. The cavity input or transmission lens tilts are adjusted to the next value we want to measure, and the measurement is repeated. To normalize our measurements, we always also record a spectrum taken at optimal alignment (at room temperature).

We fit the average off-resonant reflection values at \SI{78}{\kelvin} from five cooldowns with different input alignment settings to \cref{SI:eq:R1_2} (see \cref{SI:fig:Alignment}c,d). This produces the input lens cold optimum $\{\theta_\mathrm{in,0},\phi_\mathrm{in,0}\}^\mathrm{cold}$, which we find to be displaced from the warm optimum by (-15,23) degrees rotation on the tip-tilt mount screws.
We then proceed with the transmission lens alignment by placing the input lens at its cold optimum and performing cooldowns with five different transmission lens alignments. Before taking the \SI{78}{\kelvin} spectra, we align the back mirror in situ to its cold optimum, as is necessary to render \cref{SI:eq:T1_3} valid. We then fit \cref{SI:eq:T1_3} (see \cref{SI:fig:Alignment}e,f) to the average resonant transmission of these spectra for all five cooldowns to find the transmission lens cold optimum $\{ \theta_\mathrm{tr,0} , \phi_\mathrm{tr,0} \}$, which we find to be displaced from the warm optimum by (15,-11) degrees rotation of the tip-tilt mount screws.

To test the result of our alignment procedure, we align our cavity to the cold optimum and cool it down in our dilution fridge. We find an increase of the off-resonant reflectivity from 95\% to 100\% and an increase in resonant transmission from 50\% to 64\% (both quantities normalized to the values at warm optimum) as we cool our cavity down to $\sim\SI{30}{\milli\kelvin}$ (see \cref{fig:1_CryoCavity}d in the main paper). Note that the value quoted here are averaged over the full $\sim\SI{1}{\tera\hertz}$ spectrum, while \cref{fig:1_CryoCavity}d only shows the first \SI{100}{\giga\hertz} of these spectra, so the off-resonant reflection values do not correspond exactly.

Between the cavity alignment and the thermometry measurements presented in the main paper, 15 months passed during which the cavity was thermally cycled in our fridge nine times. To compensate for any slow drifts of the alignment, we redetermined the 'warm optimum' (the input and transmission lens optima at room temperature) before each cooldown and then changed our tilts by the warm-to-cold-optimum shifts that we found during our initial alignment. 

We should mention that this alignment only works if the cavity expands and contracts reproducibly during a cooldown. That is, it returns to its original position when warmed up. This was the case, but only after a first `settling' cooldown, during which we infer that mechanical elements overcome some stresses introduced during assembly, allowing them to remain in position in subsequent cooldowns.
Furthermore, the unmounting of an element from our experiment bracket, or even the loosening and retightening of a mounting screw, would usually lead to a loss of the calibration for that element. A new cold optimum would then have to be found by repeating the test cooldowns.

Finally, we did observe a decrease of our off-resonant reflectivity at base temperature between the first fridge cooldown and the second. Afterwards, this reflectivity remained stable at $\sim85\%$ of the optimal room-temperature value over the course of 9 cooldowns. We attribute this decrease not to input lens misalignment, which would lead to asymmetric Fano lineshapes for our cavity resonances, which we don't observe, but to a failure of the anti-reflection coatings on our input GRIN lens or fiber ferrule, which are not specified to such low temperatures. Such degradation does not appear to occur for our mirror coatings, since we do not observe a systematic broadening of the cavity linewidth.

\subsection{Selected Equipment}
\label{SI:Equipment}
$\SI{1550}{\nano\meter}$ light was created by a Toptica CTL 1550 tunable laser, intensity- and phase-modulated by Optilabs IM-1550-20-PM and iXblue MPX-LN-0.1, and sent to the experiment in a Bluefors LD400 dilution refrigerator. The local oscillator light was frequency-shifted by an iXblue MXIQER-LN-30 IQ modulator acting as a single-sideband modulator, which was driven by a Keysight MXG N5183B signal generator. For OMIT and OMIA spectra, the intensity modulator was driven by a Keysight P5004A vector network analyzer, which received its input signal from a Thorlabs RXM25AF photodetector. The spontaneous scattering signals were captured by a Thorlabs PDB570C auto-balanced photodetector and digitized by a Signalhound SM200A electrical spectrum analyzer.

The optomechanical cavity was formed by a $\SI{5}{\milli\meter}$ thick flat-flat z-cut quartz crystal from Rocky Mountain Instrument Co. between two $>99.9\%$ reflectivity mirrors from Layertec. The quartz crystal is separated from the flat front mirror by a 0.2 mm thick Teflon spacer. The back mirror with a radius of curvature of $\SI{25}{\milli\meter}$ was mounted into a JPE cryo tip-tilt piston stage driven by three CLA 2201 stick-slip piezo actuators. The in- and outcoupling GRIN lens assemblies were mounted into Thorlabs POLARIS-K05F6 mounts.

\subsection{OMIT and OMIA measurements}
\label{SI:OMITAfits}

\subsubsection{OMIT and OMIA for thermometry corrections}
\label{SI:Thermometry_OMIT_beforeafter}

\begin{figure}[b]
    \centering
    \includegraphics{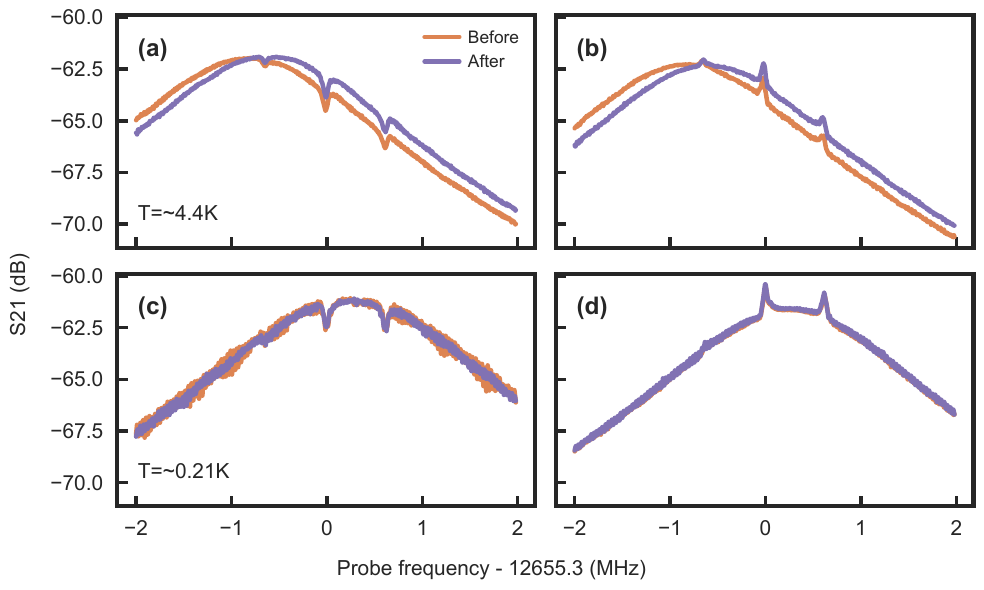}
    \caption{\textbf{OMIT and OMIA spectra before and after sideband thermometry.}
    \textbf{(a)} and \textbf{(b)} show spectra taken at around \SI{4}{\kelvin}, taken right before and after the sideband asymmetry measurements presented in \cref{fig:3_Thermometry}a-c.
    \textbf{(c)} and \textbf{(d)} show spectra taken at milliKelvin temperatures, taken right before and after the sideband asymmetry measurements presented in \cref{fig:3_Thermometry}d-f. Compared to the \SI{4}{\kelvin} measurements, more noise is visible, which comes from vibrations induced by the helium mix circulation pumps.
    }   \label{fig:SI_OMITA_shifts}
\end{figure}

As described in the main text, we perform optomechanically induced transparency and amplification measurements before and after the sideband thermometry measurements to characterize the optomechanical coupling. \cref{fig:SI_OMITA_shifts} shows the average of 20 OMIT/OMIA spectra recorded before and after the measurements in \cref{fig:3_Thermometry} of the main text. We observe that at $\sim \SI{4}{\kelvin}$, a slight shift of the optical mode spacing occurred during the measurement, which we attribute to a thermal expansion of the experiment. At \si{\milli\kelvin} temperatures, no significant shift is visible. We attribute this to the fact that the helium circulation still provides active cooling to the experiment at \si{\milli\kelvin}, whereas it is turned off at \SI{4}{\kelvin}. 

\subsubsection{Fitting OMIT and OMIA spectra}

We fit the averaged OMIT/OMIA spectra to extract the relevant parameters, such as the mechanical peak positions with respect to the optical resonance, the optical and mechanical linewidths and the optomechanical coupling rates. 
The transmitted probe tone intensity spectrum for an optical cavity coupled to a single mechanical mode is given by \cite{Kharel2022}
\begin{align}
    \label{eqSIOMITA:I_transmitted}
    \abs{I_T(\Omega)}^2 = \abs{ A_0 \frac{\frac{\kappa}{2}}{\frac{\kappa}{2} - i ( \Omega - \Delta_{21}) \pm \frac{\abs{g}^2}{\frac{\Gamma_m}{2} - i (\Omega - \Omega_m)}} }^2,
\end{align}
where $\Delta_{21}=\omega_2-\omega_1$ is the frequency detuning between the two optical modes, $\kappa$ is the optical linewidth, $g$ is the cavity-enhanced coupling rate, $\Omega_m$ is the mechanical frequency, $\Gamma_m$ is the intrinsic mechanical linewidth and $A_0$ is the transmission amplitude. We have assumed our pump laser to be resonant with one of the optical modes. The effective mechanical linewidth $\Gamma_{m,\mathrm{eff}}$, which includes optomechanical backaction and is shown in \cref{fig:2_OMITA}d,e of the main paper, is calculated using \cref{eqSIderiv:Gamma_mrb_definition}.
The last term in the denominator enters with a plus (minus) and causes a narrow dip (peak) on the broad optical resonance when the pump is locked to the low (high) frequency optical mode at $\omega_1$ ($\omega_2$), corresponding to the case of OMIT (OMIA).

\begin{figure}[b]
    \centering
    \includegraphics{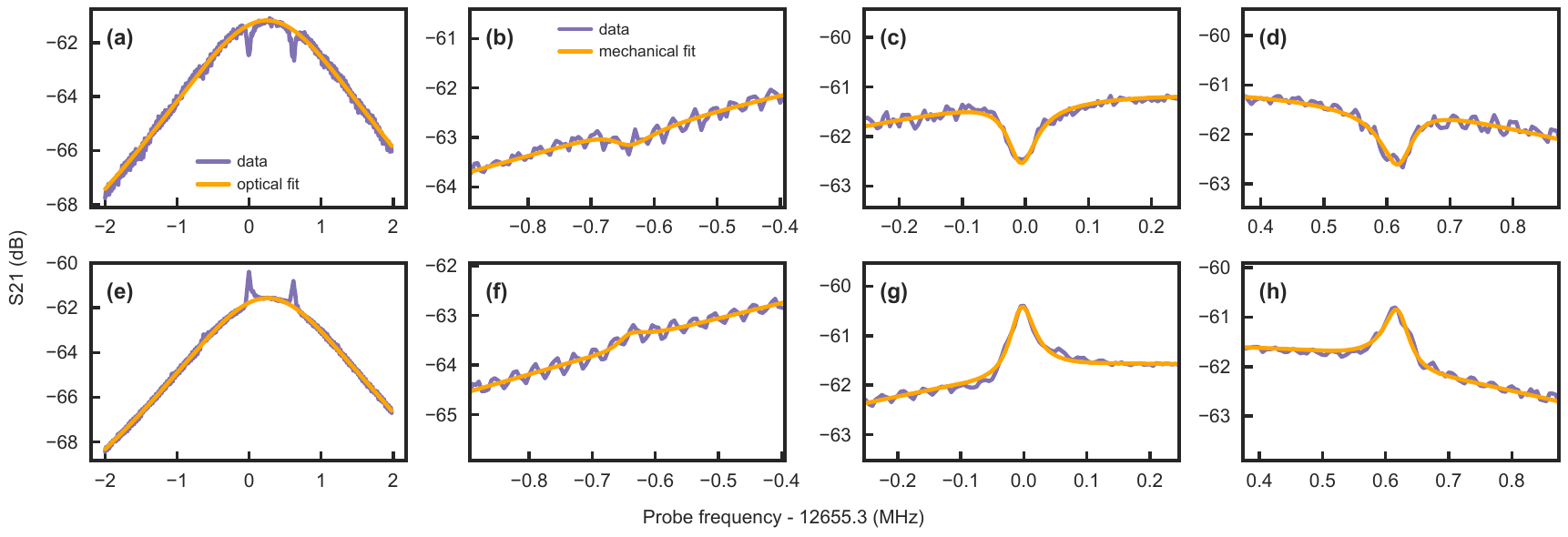}
    \caption{\textbf{Fitting OMIT and OMIA spectra.}
    \textbf{(a-d)}~OMIT and \textbf{(e-h)}~OMIA spectra taken after the thermometry measurement at $\sim\SI{200}{\milli\kelvin}$ presented in \cref{fig:3_Thermometry}d-f. \textbf{(a)}~and \textbf{(e)}~show the full spectra with the fit to extract the optical resonance parameters. \textbf{(b-d)}~and \textbf{(f-h)}~show closeups of the regions used to fit the mechanical resonances individually.
    }   \label{fig:SI_OMITA_fits}
\end{figure}

For each OMIT/OMIA measurement, the recorded traces are preprocessed in multiple stages and several fits are preformed to extract the optical resonance or the mechanical resonances. First, a simple Lorentzian lineshape (the 'optical fit') is fitted to the broad optical mode by ignoring the data points near the mechanical peaks (see \cref{fig:SI_OMITA_fits}a,e).
With the optical mode parameters $\Delta_{21}, \kappa$ and $A_0$ fixed by the optical fit, we then fit the region around each mechanical peak (dip) individually using \cref{eqSIOMITA:I_transmitted}, as shown in \cref{fig:SI_OMITA_fits}b-d,f-h (the 'mechanical fits').
The uncertainties on the fit parameters are propagated using linear error propagation when using them for the signal corrections as described in \cref{SI:Thermometry_corrections}.

There are two further sources of errors on the position of the optical resonance which are not captured by a single fit. First, any change in parameters between before and after the thermometry measurement, as discussed in \cref{SI:Thermometry_OMIT_beforeafter}, is taken into account by taking the average parameter value and adding an error of half the change to either side. Second, at \si{\milli\kelvin}, there are sinusoidal oscillations of the optical resonance spacing $\Delta_{21}$ due to noise from the turbo pumps that circulate the helium mix (see \cref{fig:SI_OMITA_shifts}c,d. These oscillations mostly cancel out when averaging multiple OMIT/OMIA spectra, but are clearly visible in a single trace. From a single OMIT trace taken for the \si{\milli\kelvin} measurement in \cref{fig:3_Thermometry}d-f, we estimate the amplitude of the signal mode frequency fluctuations by using the amplitude of the oscillations of the transmitted intensity and the slope of the trace at this position. We find a conservative estimate for the root mean square amplitude of the frequency fluctuations of $\SI{0.266}{\mega\hertz}$, which we add to the uncertainty on $\Delta_{21}$ for every OMIT/OMIA measurement taken at \si{\milli\kelvin} temperatures. This source of error is roughly equal to the total contributions by other error sources when calculating the error bars for the thermal mode occupations. 

\subsection{Thermometry signals in a two-mode optical cavity}
\label{SI:Thermometry_theory}

\subsubsection{Optical cavity thermometry output signal}
\label{SI:Thermometry_cavity_signal}

In the following we will derive the signal observed on the electrical spectrum analyzer (ESA, see \cref{fig:2_OMITA}a) created by optomechanical up- or downconversion when pumping either the red or the blue mode with a strong classical pump tone. We will limit the derivation to the interaction with a single mechanical mode $\hat{b}$ for notational brevity, even though in the experiment we observe optomechanical coupling to multiple mechanical modes $\hat{b}_m$. However, since the mechanical modes are well separated in frequency, we can treat them as independent. Although the different mechanical modes exhibit similar linewidths, their single photon coupling rates $g_0$ are strongly modulated, as found previously by Kharel \textit{et. al.} \cite{Kharel2019}.

The system model consists of two optical modes $\hat{a}_{1/2}$ and a mechanical mode $\hat{b}$ coupled with rate $g_0$. The optical modes have linewidths $\kappa_{1/2} = \kappa_{1/2}^\mathrm{int} + \kappa_{1/2}^\mathrm{ext1} + \kappa_{1/2}^\mathrm{ext2}$, with $\kappa_{1/2}^\mathrm{int}$ the internal losses, $\kappa_{1/2}^\mathrm{ext1}$ the rate with which light from $\hat{a}_{1/2,in}^\mathrm{ext1}$ is coupled into the cavity mode, and $\kappa_{1/2}^\mathrm{ext2}$ the rate with which light exits the cavity into the propagating mode $\hat{a}_{1/2,out}^\mathrm{ext2}$ leading to the detector. A strong classical pump tone with amplitude $\alpha_{p,in}$ and frequency $\omega_p$ drives one of the two modes, designated $\hat{a}_{1/2}$ for the red (blue) pumping case. The full Hamiltonian for this system is
\begin{align}
    \label{eqSIderiv:fullH_start}
    \hat{H} = \hbar \omega_1 \hat{a}_1^\dagger \hat{a}_1 
    + \hbar \omega_2 \hat{a}_2^\dagger \hat{a}_2
    + \hbar \Omega_m \hat{b}^\dagger \hat{b}
    - \hbar g_0 \qty( \hat{a}_1 \hat{a}_2^\dagger \hat{b} + \hat{a}_1^\dagger \hat{a}_2 \hat{b}^\dagger )
    + i\hbar \sqrt{\kappa_{1/2}^\mathrm{ext1}} \alpha_{p,in} \qty( \hat{a}_{1/2}^\dagger e^{-i\omega_p t} - \hat{a}_{1/2} e^{i\omega_p t} ).
\end{align}
The Langevin equation of motion for the pump mode is thus
\begin{align}
    \label{eqSIderiv:LE_pump}
    \Dot{\hat{a}}_{1/2} = -i \omega_{1/2} \hat{a}_{1/2} + \sqrt{\kappa_{1/2}^\mathrm{ext1}} \alpha_{p,in} e^{-i\omega_p t} - \frac{\kappa_{1/2}}{2} \hat{a}_{1/2},
\end{align}
where we have assumed weak single photon coupling to neglect the term $i g_0 \hat{a}_{2/1} \hat{b}$. Neglecting also quantum fluctuations of the pump mode amplitude, i.e. replacing $\hat{a}_{1/2} \rightarrow \Bar{a}_{1/2}$, we solve for the classical pump mode amplitude
\begin{align}
    \Bar{a}_{1/2} &= \frac{\sqrt{\kappa_{1/2}^\mathrm{ext1}}}{\frac{\kappa_{1/2}}{2} - i \Delta_{p,1/2}} \alpha_{p,in} e^{-i\omega_p t} \nonumber \\
    &\equiv \alpha_p^\mathrm{cav} e^{-i\omega_p t},
\end{align}
where $\Delta_{p,1/2} = \omega_p - \omega_{1/2}$ and $\alpha_p^\mathrm{cav}$ is the intra-cavity amplitude of the pump mode. Thus we identify $\abs{\alpha_p^\mathrm{cav}}^2 = N_{1/2}$ as the number of intra-cavity photons of the pump mode. For simplicity, we define the time $t$ to absorb the complex phase of $\alpha_p^\mathrm{cav}$ such that $\alpha_p^\mathrm{cav}=\sqrt{N_{1/2}}$ is real. We insert $\Bar{a}_{1/2}$ for $\hat{a}_{1/2}$ in \cref{eqSIderiv:fullH_start} to obtain
\begin{align}
    \label{eqSIderiv:fullH_pumpelim}
    \hat{H} = \hbar \omega_{1/2} N_{1/2}
    + \hbar \omega_{2/1} \hat{a}_{2/1}^\dagger \hat{a}_{2/1}
    + \hbar \Omega_m \hat{b}^\dagger \hat{b}
    + \begin{cases}
            - \hbar g_r \qty( \hat{a}_2^\dagger \hat{b} e^{-i\omega_p t} + \hat{a}_2 \hat{b}^\dagger e^{i\omega_p t} ) & \text{for red pumping} \\
            - \hbar g_b \qty( \hat{a}_1 \hat{b} e^{i\omega_p t} + \hat{a}_1^\dagger \hat{b}^\dagger e^{-i\omega_p t} ) & \text{for blue pumping.}
        \end{cases}
\end{align}
where we defined the cavity-enhanced coupling rates $g_{r/b} = g_0 \sqrt{N_{1/2}}$ for the red (blue) pumping case. Going to the rotating frame with respect to $\hbar \omega_p \hat{a}_{2/1}^\dagger \hat{a}_{2/1}$ and assuming that the pump beam is on resonance, i.e. $\omega_p = \omega_{1/2}$ yields
\begin{align}
    \label{eqSIderiv:fullH_rot}
    \hat{\widetilde{H}} = \hbar \omega_{1/2} N_{1/2}
    \pm \hbar \Delta_{21} \hat{a}_{2/1}^\dagger \hat{a}_{2/1}
    + \hbar \Omega_m \hat{b}^\dagger \hat{b}
    + \begin{cases}
            - \hbar g_r \qty( \hat{a}_2^\dagger \hat{b} + \hat{a}_2 \hat{b}^\dagger ) & \text{for red pumping} \\
            - \hbar g_b \qty( \hat{a}_1 \hat{b} + \hat{a}_1^\dagger \hat{b}^\dagger ) & \text{for blue pumping,}
        \end{cases}
\end{align}
where $\Delta_{21} = \omega_2 - \omega_1 > 0$.
This leads to the following Langevin equations of motion for the signal mode $\hat{a}_{2/1}$ and the mechanical mode $\hat{b}$:
\begin{align}
    \label{eqSIderiv:LE_signal_and_mech}
    \Dot{\hat{a}}_{2/1} &= \mp i \Delta_{21} \hat{a}_{2/1} - \frac{\kappa_{2/1}}{2} \hat{a}_{2/1} + \sqrt{\kappa_{2/1}^\mathrm{ext1}} \hat{a}_{2/1,in}^\mathrm{ext1} + \sqrt{\kappa_{2/1}^\mathrm{ext2}} \hat{a}_{2/1,in}^\mathrm{ext2} + \sqrt{\kappa_{2/1}^\mathrm{int}} \hat{a}_{2/1,in}^\mathrm{int} +
    C_{\hat{a}_{2/1}}^{r/b} \\
    \Dot{\hat{b}} &= - i \Omega_m \hat{b} - \frac{\Gamma_m}{2} \hat{b} + \sqrt{\Gamma_m} \hat{b}_\mathrm{in} +
    C_{\hat{b}}^{r/b}
\end{align}
where we consider the mechanical mode only coupled to one single loss channel, and defined the coupling terms
\begin{align}
    C_{\hat{a}_{2/1}}^{r/b} &=
    \begin{cases}
        i g_r \hat{b} & \text{for red pumping} \\
        i g_b \hat{b}^\dagger & \text{for blue pumping.}
    \end{cases} \\
    C_{\hat{b}}^{r/b} &=
    \begin{cases}
        i g_r \hat{a}_2 & \text{for red pumping} \\
        i g_b \hat{a}_1^\dagger & \text{for blue pumping.}
    \end{cases}
\end{align}
Applying the Fourier transformation defined as $A(\omega) = \int_{-\infty}^\infty \dd t e^{i \omega t} A(t)$, using $\hat{A}^\dagger(\omega) = \qty(\hat{A}(-\omega))^\dagger$ and defining the optical noise input to the signal mode
\begin{align}
    \hat{\xi}_{2/1}(\omega) = \sqrt{\kappa_{2/1}^\mathrm{ext1}} \hat{a}_{2/1,in}^\mathrm{ext1}(\omega) + \sqrt{\kappa_{2/1}^\mathrm{ext2}} \hat{a}_{2/1,in}^\mathrm{ext2}(\omega) + \sqrt{\kappa_{2/1}^\mathrm{int}} \hat{a}_{2/1,in}^\mathrm{int}(\omega)
\end{align}
yields
\begin{align}
    \hat{a}_{2/1}(\omega) &= \frac{C_{\hat{a}_{2/1}}^{r/b}(\omega) + \hat{\xi}_{2/1}(\omega)}{\frac{\kappa_{2/1}}{2} - i \qty( \omega \mp \Delta_{21} )} \label{eqSIderiv:optsignalmode_intermediate} \\
    \hat{b}(\omega) &= \frac{C_{\hat{b}}^{r/b}(\omega) + \sqrt{\Gamma_m} \hat{b}_\mathrm{in}(\omega)}{\frac{\Gamma_m}{2} - i \qty( \omega - \Omega_m )} \\
    \Rightarrow \hat{b}(\omega) &= \frac{1}{\frac{\Gamma_m}{2} - i \qty(\omega - \Omega_m) \pm i \underbrace{\frac{g_{r/b}^2}{\qty(\omega - \Delta_{21}) + i\frac{\kappa_{2/1}}{2}}}_{\equiv \Sigma^{r/b}(\omega)}} \qty[ \sqrt{\Gamma_m} \hat{b}_\mathrm{in}(\omega) + \frac{C_{\hat{\xi}_{2/1}}(\omega)}{\frac{\kappa_{2/1}}{2} - i \qty(\omega - \Delta_{21})} ] \nonumber \\
    &= \frac{1}{\frac{\Gamma_m^{r/b}}{2} - i \qty(\omega - \Omega_m^{r/b})} \qty[ \sqrt{\Gamma_m} \hat{b}_\mathrm{in}(\omega) + \frac{C_{\hat{\xi}_{2/1}}(\omega)}{\frac{\kappa_{2/1}}{2} - i \qty(\omega - \Delta_{21})} ] \label{eqSIderiv:mech_final}
\end{align}
where
\begin{align}
    C_{\hat{\xi}_{2/1}}(\omega) &=
    \begin{cases}
        i g_r \hat{\xi}_2(\omega) & \text{for red pumping} \\
        i g_b \hat{\xi}_1^\dagger(\omega) & \text{for blue pumping}
    \end{cases} \\
\end{align}
and we define the effective mechanical frequency and linewidth, modified by optomechanical backaction, as
\begin{align}
    \Omega_m^{r/b} &= \Omega_m \pm \delta\Omega_m^{r/b} \\
    \Gamma_m^{r/b} &= \Gamma_m \pm \delta\Gamma_m^{r/b} \label{eqSIderiv:Gamma_mrb_definition} \\
    \Sigma^{r/b}(\omega) &= \delta\Omega_m^{r/b}(\omega) - i \frac{\delta\Gamma_m^{r/b}(\omega)}{2} \\
    \Rightarrow \delta\Omega_m^{r/b}(\omega) &= \Re{\Sigma^{r/b}(\omega)} = \frac{g_{r/b}^2 \qty(\omega - \Delta_{21})}{\qty(\omega - \Delta_{21})^2 + \qty(\frac{\kappa_{2/1}}{2})^2} \\
    \Rightarrow \delta\Gamma_m^{r/b}(\omega) &= -2 \Im{\Sigma^{r/b}(\omega)} = \frac{g_{r/b}^2 \kappa_{2/1}}{\qty(\omega - \Delta_{21})^2 + \qty(\frac{\kappa_{2/1}}{2})^2}. \label{eqSIderiv:deltaGamma_mrb_definition}
\end{align}

Inserting \cref{eqSIderiv:mech_final} back into \cref{eqSIderiv:optsignalmode_intermediate} yields
\begin{align}
    \hat{a}_{2/1}(\omega) &= \frac{1}{\frac{\kappa_{2/1}}{2} - i \qty( \omega \mp \Delta_{21} )} \qty[ \frac{\sqrt{\Gamma_m}}{\frac{\Gamma_m^{r/b}}{2} - i \qty(\omega \mp \Omega_m^{r/b})} C_{\hat{b}_\mathrm{in}}^{r/b} + \qty( 1 \mp \frac{g_{r/b}^2}{\qty(\frac{\kappa_{2/1}}{2} - i \qty( \omega \mp \Delta_{21} ))\qty(\frac{\Gamma_m^{r/b}}{2} - i \qty(\omega \mp \Omega_m^{r/b}))} ) \hat{\xi}_{2/1}(\omega)], \label{eqSIderiv:optsignalmode_final}
\end{align}
where
\begin{align}
    C_{\hat{b}_\mathrm{in}}^{r/b}(\omega) &=
    \begin{cases}
        i g_r \hat{b}_\mathrm{in}(\omega) & \text{for red pumping} \\
        i g_b \hat{b}_\mathrm{in}^\dagger(\omega) & \text{for blue pumping.}
    \end{cases}
\end{align}
Note that by taking the Hermitian conjugate of $\hat{b}(\omega)$ in the blue pumping case, the frequency argument of $\Omega_m^b$ and $\Gamma_m^b$ flipped its sign, meaning
\begin{align}
    \Omega_m^{r/b} &\equiv
    \begin{cases}
        \Omega_m^r(\omega) & \text{for red pumping} \\
        \Omega_m^b(-\omega) & \text{for blue pumping}
    \end{cases} \label{eqSIderiv:Omega_mb_signflip} \\
    \Gamma_m^{r/b} &\equiv
    \begin{cases}
        \Gamma_m^r(\omega) & \text{for red pumping} \\
        \Gamma_m^b(-\omega) & \text{for blue pumping.}
    \end{cases} \label{eqSIderiv:Gamma_mb_signflip}
\end{align}

Finally, \cref{eqSIderiv:optsignalmode_final} together with $\hat{a}_{2/1,out}^\mathrm{ext2}(\omega) = \hat{a}_{2/1,in}^\mathrm{ext2}(\omega) - \sqrt{\kappa_{2/1}^\mathrm{ext2}} \hat{a}_{2/1}(\omega)$ then yields the signal in the optical output mode $\hat{a}_{2/1,out}^\mathrm{ext2}(\omega)$. We can write the relations between all mechanical and optical input and output modes in terms of a scattering matrix:
\begin{equation}
\label{eqSIderiv:Scattering_relation}
    \begin{pmatrix}
        \hat{a}_{2/1,out}^\mathrm{ext1}(\omega) \\
        \hat{a}_{2/1,out}^\mathrm{ext2}(\omega) \\
        \hat{B}_{out}^{r/b}(\omega) \\
        \hat{a}_{2/1,out}^\mathrm{int}(\omega)
    \end{pmatrix}
    = \underline{\mathcal{S}}^{r/b}(\omega)
    \begin{pmatrix}
        \hat{a}_{2/1,in}^\mathrm{ext1}(\omega) \\
        \hat{a}_{2/1,in}^\mathrm{ext2}(\omega) \\
        \hat{B}_\mathrm{in}^{r/b}(\omega) \\
        \hat{a}_{2/1,in}^\mathrm{int}(\omega)
    \end{pmatrix}
\end{equation}
where
\begin{align}
    \hat{B}_{in/out}^{r/b}(\omega) = 
    \begin{cases}
        \hat{b}_{in/out}(\omega) & \text{for red pumping} \\
        \hat{b}_{in/out}^\dagger(\omega) & \text{for blue pumping.}
    \end{cases}
\end{align}
For this experiment, the relevant scattering matrix element is $\mathcal{S}_{23}^{r/b}(\omega)$, connecting $\hat{a}_{2/1,out}^\mathrm{ext2}(\omega)$ and $\hat{B}_\mathrm{in}^{r/b}(\omega)$, which is
\begin{align}
    \mathcal{S}_{23}^{r/b}(\omega) &= - \frac{i g_{r/b} \sqrt{\Gamma_m} \sqrt{\kappa_{2/1}^\mathrm{ext2}} }{\qty(\frac{\kappa_{2/1}}{2} - i \qty( \omega \mp \Delta_{21} ))\qty(\frac{\Gamma_m^{r/b}}{2} - i \qty(\omega \mp \Omega_m^{r/b}))} \label{eqSIderiv:S23} \\
    \Rightarrow \quad \abs{\mathcal{S}_{23}^{r/b}(\omega)}^2 &= g_{r/b}^2 \frac{ \kappa_{2/1}^\mathrm{ext2} }{\qty(\frac{\kappa_{2/1}}{2})^2 + \qty( \omega \mp \Delta_{21} )^2 } \frac{\Gamma_m}{\qty(\frac{\Gamma_m^{r/b}}{2})^2 + \qty(\omega \mp \Omega_m^{r/b})^2}
    \label{eqSIderiv:S23_mag2}
\end{align}

Due to energy conservation, the magnitude squared of the four scattering matrix elements $\mathcal{S}_{2j}^{r/b}$ add up to one, although in the blue pumping case, $\abs{\mathcal{S}_{23}^b(\omega)}^2$ enters with a minus sign (recall the different definitions of $\hat{B}_{in/out}^{r/b}(\omega)$ in \cref{eqSIderiv:Scattering_relation}).

For later reference, we now prove the relation
\begin{align}
\label{eqSIderiv:Sblue_conservation}
    - \abs{\mathcal{S}_{23}^b(\omega)}^2 + \sum_{j=1,2,4}\abs{\mathcal{S}_{2j}^b(\omega)}^2 &= 1.
\end{align}
For this, we write down the scattering matrix elements for the blue pumping case:
\begin{align}
    \mathcal{S}_{21}^b(\omega)
    &= -\frac{\sqrt{\kappa_1^\mathrm{ext1}}\sqrt{\kappa_1^\mathrm{ext2}}}{\frac{\kappa_1}{2} - i \qty( \omega + \Delta_{21} )} \qty( 1 + \frac{g_b^2}{\qty(\frac{\kappa_1}{2} - i \qty( \omega + \Delta_{21} )) \qty(\frac{\Gamma_m^b}{2} - i \qty(\omega + \Omega_m^b))} ) \nonumber \\
    &\equiv - \sqrt{\kappa_1^\mathrm{ext1}} D(\omega) \qty( 1 + E(\omega) ) \label{eqSIderiv:S21b} \\
    \mathcal{S}_{22}^b(\omega)
    &= 1 - \frac{\kappa_1^\mathrm{ext2}}{\frac{\kappa_1}{2} - i \qty( \omega + \Delta_{21} )} \qty( 1 + \frac{g_b^2}{\qty(\frac{\kappa_1}{2} - i \qty( \omega + \Delta_{21} )) \qty(\frac{\Gamma_m^b}{2} - i \qty(\omega + \Omega_m^b))} ) \nonumber \\
    &\equiv 1 - \sqrt{\kappa_1^\mathrm{ext2}} D(\omega) \qty( 1 + E(\omega) ) \label{eqSIderiv:S22b} \\
    \mathcal{S}_{23}^b(\omega)
    &= -\frac{i g_b \sqrt{\kappa_1^\mathrm{ext2}} \sqrt{\Gamma_m}}{\qty(\frac{\kappa_1}{2} - i \qty( \omega + \Delta_{21} )) \qty(\frac{\Gamma_m^b}{2} - i \qty(\omega + \Omega_m^b))} \nonumber \\
    &\equiv \sqrt{\kappa_1^\mathrm{ext2}} \sqrt{\Gamma_m} \frac{E(\omega)}{i g_b} \label{eqSIderiv:S23b} \\
    \mathcal{S}_{24}^b(\omega)
    &= -\frac{\sqrt{\kappa_1^\mathrm{int}}\sqrt{\kappa_1^\mathrm{ext2}}}{\frac{\kappa_1}{2} - i \qty( \omega + \Delta_{21} )} \qty( 1 + \frac{g_b^2}{\qty(\frac{\kappa_1}{2} - i \qty( \omega + \Delta_{21} )) \qty(\frac{\Gamma_m^b}{2} - i \qty(\omega + \Omega_m^b))} ) \nonumber \\
    &\equiv - \sqrt{\kappa_1^\mathrm{int}} D(\omega) \qty( 1 + E(\omega) ), \label{eqSIderiv:S24b}
\end{align}
where
\begin{align}
    D(\omega) &= \frac{\sqrt{\kappa_1^\mathrm{ext2}}}{\frac{\kappa_1}{2} - i \qty( \omega + \Delta_{21} )} \\
    E(\omega) &= \frac{g_b^2}{\qty(\frac{\kappa_1}{2} - i \qty( \omega + \Delta_{21} )) \qty(\frac{\Gamma_m^b}{2} - i \qty(\omega + \Omega_m^b))}.
\end{align}
So the magnitude squared of the scattering matrix elements are
\begin{align}
    \abs{\mathcal{S}_{21}^b(\omega)}^2
    &=  \kappa_1^\mathrm{ext1} \abs{D(\omega)}^2 \abs{ 1 + E(\omega) }^2 \label{eqSIderiv:magS21b} \\
    \abs{\mathcal{S}_{22}^b(\omega)}^2
    &=  1 + \kappa_1^\mathrm{ext2} \abs{D(\omega)}^2 \abs{ 1 + E(\omega) }^2 \nonumber \\
    &\quad - \abs{D(\omega)}^2 \qty ( \frac{\sqrt{\kappa_1^\mathrm{ext2}}}{\qty(D(\omega))^\dagger} \qty( 1 + E(\omega) ) + \frac{\sqrt{\kappa_1^\mathrm{ext2}}}{D(\omega)} \qty( 1 + \qty(E(\omega))^\dagger ) ) \label{eqSIderiv:magS22b} \\
    \abs{\mathcal{S}_{23}^b(\omega)}^2
    &=  \frac{\kappa_1^\mathrm{ext2} \Gamma_m}{g_b^2} \abs{E(\omega)}^2 \label{eqSIderiv:magS23b} \\
    \abs{\mathcal{S}_{24}^b(\omega)}^2
    &=  \kappa_1^\mathrm{int} \abs{D(\omega)}^2 \abs{1 + E(\omega) }^2. \label{eqSIderiv:magS24b}
\end{align}
Here, we identify
\begin{align}
    \frac{\sqrt{\kappa_1^\mathrm{ext2}}}{\qty(D(\omega))^\dagger} &= \frac{\kappa_1}{2} + i \qty( \omega + \Delta_{21} ) \\
    \frac{\sqrt{\kappa_1^\mathrm{ext2}}}{D(\omega)} &= \frac{\kappa_1}{2} - i \qty( \omega + \Delta_{21} )
\end{align}
and therefore the last term in \cref{eqSIderiv:magS22b} becomes
\begin{align}
    &\quad \qty(\frac{\kappa_1}{2} + i \qty( \omega + \Delta_{21} )) \qty( 1 + E(\omega) ) + \qty(\frac{\kappa_1}{2} - i \qty( \omega + \Delta_{21} )) \qty( 1 + \qty(E(\omega))^\dagger ) \nonumber \\
    &= \frac{\kappa_1}{2} \qty( 1 + E(\omega) + 1 + \qty(E(\omega))^\dagger ) + i \qty( \omega + \Delta_{21} ) \qty( 1 + E(\omega) - 1 - \qty(E(\omega))^\dagger ) \nonumber \\
    &= \kappa_1 \qty( 1 + \Re{E(\omega)}) - 2\qty( \omega + \Delta_{21} ) \Im{E(\omega)}.
\end{align}
Using $\kappa_1 = \kappa_1^\mathrm{ext1} + \kappa_1^\mathrm{ext2} + \kappa_1^\mathrm{int}$, \cref{eqSIderiv:Sblue_conservation} thus becomes
\begin{align}
    1 &= 1 - \frac{\kappa_1^\mathrm{ext2} \Gamma_m}{g_b^2} \abs{E(\omega)}^2 + \abs{D(\omega)}^2 \qty( \kappa_1 \abs{ 1 + E(\omega) }^2 - \kappa_1 \qty( 1 + \Re{E(\omega)}) + 2\qty( \omega + \Delta_{21} ) \Im{E(\omega)} ) \\
    \Rightarrow \quad 0 &= -\frac{\kappa_1^\mathrm{ext2} \Gamma_m}{g_b^2} \frac{\abs{E(\omega)}^2}{\abs{D(\omega)}^2} + \kappa_1 \qty( \Re{E(\omega)} + \abs{E(\omega)}^2 ) + 2\qty( \omega + \Delta_{21} ) \Im{E(\omega)} ). \\
    \Rightarrow \quad 0 &= - \frac{g_b^2 \Gamma_m}{\qty(\frac{\Gamma_m^b}{2})^2 + \qty(\omega + \Omega_m^b)^2} + \frac{1}{\qty(\qty(\frac{\kappa_1}{2})^2 + \qty( \omega + \Delta_{21} )^2) \qty(\qty(\frac{\Gamma_m^b}{2})^2 + \qty(\omega + \Omega_m^b)^2)} \nonumber \\
    &\quad \times \qty( \kappa_1 g_b^2 \qty( \frac{\kappa_1}{2} \frac{\Gamma_m^b}{2} - \qty( \omega + \Delta_{21} ) \qty(\omega + \Omega_m^b) )  + \kappa_1 g_b^4 + 2\qty( \omega + \Delta_{21} ) g_b^2 \qty( \frac{\kappa_1}{2} \qty(\omega + \Omega_m^b) + \frac{\Gamma_m^b}{2} \qty( \omega + \Delta_{21} ) ) ).
\end{align}
Dividing by $\frac{g_b^2}{\qty(\frac{\Gamma_m^b}{2})^2 + \qty(\omega + \Omega_m^b)^2}$, adding $\Gamma_m$ and subsequently multiplying by $\qty(\frac{\kappa_1}{2})^2 + \qty( \omega + \Delta_{21} )^2$ yields
\begin{align}
    \Gamma_m \qty(\qty(\frac{\kappa_1}{2})^2 + \qty( \omega + \Delta_{21} )^2) &= \kappa_1 \qty( \frac{\kappa_1}{2} \frac{\Gamma_m^b}{2} - \qty( \omega + \Delta_{21} ) \qty(\omega + \Omega_m^b) ) + \kappa_1 g_b^2 \nonumber \\
    &\quad + 2\qty( \omega + \Delta_{21} ) \qty( \frac{\kappa_1}{2} \qty(\omega + \Omega_m^b) + \frac{\Gamma_m^b}{2} \qty( \omega + \Delta_{21} ) ) \\
    &= \kappa_1 \frac{\kappa_1}{2} \frac{\Gamma_m^b}{2}  + \kappa_1 g_b^2 + \Gamma_m^b \qty( \omega + \Delta_{21} )^2 \nonumber \\
    \Rightarrow \quad \Gamma_m - \Gamma_m^b &= \frac{\kappa_1 g_b^2}{\qty(\frac{\kappa_1}{2})^2 + \qty( \omega + \Delta_{21} )^2}. \label{eqSIderiv:Sblue_conservation_final_equation}
\end{align}
By comparing the left side of this equation with \cref{eqSIderiv:Gamma_mrb_definition,eqSIderiv:Gamma_mb_signflip}, we identify $\Gamma_m - \Gamma_m^b$ as $\delta\Gamma_m^b(-\omega)$. inserting $-\omega$ into \cref{eqSIderiv:deltaGamma_mrb_definition}, we find
\begin{align}
    \delta\Gamma_m^b(-\omega) &= \frac{g_{r/b}^2 \kappa_{2/1}}{\qty(-\omega - \Delta_{21})^2 + \qty(\frac{\kappa_{2/1}}{2})^2} \nonumber \\
    &= \frac{g_{r/b}^2 \kappa_{2/1}}{\qty(\omega + \Delta_{21})^2 + \qty(\frac{\kappa_{2/1}}{2})^2},
\end{align}
which is just the right side of \cref{eqSIderiv:Sblue_conservation_final_equation}, so \cref{eqSIderiv:Sblue_conservation} holds.

\subsubsection{Output voltage of the balanced detector}

The output mode of the optical cavity is collected into the single-mode fiber that guides the signal to the detector with an amplitude collection efficiency of $\sqrt{\eta}$. Before the detector, the signal is split by a beamsplitter with intensity transmission $T$ into two paths. From the second input port of the beamsplitter, a second mode is added, in this case the strong local oscillator with amplitude $\hat{a}_\mathrm{LO} = \alpha_\mathrm{LO} e^{-i \Delta_\mathrm{LO}^{r/b} t}$ in the rotating frame of the pump laser, where $\Delta_\mathrm{LO} = \omega_\mathrm{LO} - \omega_p \rightarrow \Delta_\mathrm{LO}^{r/b} = \omega_\mathrm{LO} - \omega_{1/2}$. In our case, the local oscillator is tuned close to the signal frequency such that $\Delta_\mathrm{LO}^{r/b} / 2\pi \sim \pm \SI{12.5}{\giga\hertz}$. The signal in the mode impinging onto one of the detectors is thus
\begin{align}
\label{eqSIderiv:detector_mode}
    \hat{a}_{det1} = \sqrt{1-T} \hat{a}_\mathrm{LO} + \sqrt{T}\sqrt{\eta}\hat{a}_{2/1,out}^\mathrm{ext2}
\end{align}

In the following, we will denote $\hat{a}_{2/1,out}^\mathrm{ext2}$ as $\hat{a}_s$, $\omega_{2/1}$ as $\omega_s$ and $\Delta_\mathrm{LO}^{r/b}$ as $\Delta_\mathrm{LO}$ for clarity, until it becomes necessary again to distinguish the red/blue pumping cases. We will first derive the voltage $V_1$ of one of the photodiodes in our auto-balanced detector, and later consider the effect of substracting the photocurrents from both photodiodes. The voltage produced by a photodiode upon light absorption is given by
\begin{align}
    V_1 = G \hbar \omega_{det} \hat{a}_{det1}^\dagger \hat{a}_{det1},
    \label{eqSIderiv:V1(t)}
\end{align}
where $G$ is the photodiode gain in $\si{\volt}/\si{\watt}$, $\hbar \omega_{det}$ is the photon energy and $\hat{a}_{det1}^\dagger \hat{a}_{det1}$ is the photon flux in the detector mode. Fourier transforming \cref{eqSIderiv:V1(t)} and using the convolution theorem leads to
\begin{align}
    V_1(\omega) &= G \hbar \omega_{det} \int_{-\infty}^\infty \hat{a}_{det1}^\dagger(\omega') \hat{a}_{det1}(\omega-\omega') \dd \omega' \\
    &= G\hbar \int_{-\infty}^\infty \dd \omega' \biggl[ 
    (1-T)\omega_\mathrm{LO} \hat{a}_\mathrm{LO}^\dagger(\omega') \hat{a}_\mathrm{LO}(\omega - \omega') \nonumber \\
    &\quad+ T \eta \omega_s \hat{a}_s^\dagger(\omega') \hat{a}_s(\omega - \omega') \nonumber \\
    &\quad+ \sqrt{T(1-T) \eta \omega_\mathrm{LO} \omega_s} \qty( \hat{a}_\mathrm{LO}^\dagger(\omega') \hat{a}_s(\omega - \omega') + \hat{a}_s^\dagger(\omega') \hat{a}_\mathrm{LO}(\omega - \omega')) \biggr] \\
    &= G \hbar \int_{-\infty}^\infty \dd \omega' \biggl[ 
    2\pi (1-T)\omega_\mathrm{LO} \abs{\alpha_\mathrm{LO}}^2 \delta(\omega'+ \Delta_\mathrm{LO}) \delta(\omega - \omega'- \Delta_\mathrm{LO}) \nonumber \\
    &\quad+ T \eta \omega_s \hat{a}_s^\dagger(\omega') \hat{a}_s(\omega - \omega') \nonumber \\
    &\quad+ \sqrt{2 \pi T(1-T) \eta \omega_\mathrm{LO} \omega_s} \qty( \alpha_\mathrm{LO}^\dagger \delta(\omega'+ \Delta_\mathrm{LO}) \hat{a}_s(\omega - \omega') + \alpha_\mathrm{LO} \hat{a}_s^\dagger(\omega') \delta(\omega - \omega'- \Delta_\mathrm{LO})) \biggr] \\
    &= G \hbar\biggl[ 
    2\pi (1-T)\omega_\mathrm{LO} \abs{\alpha_\mathrm{LO}}^2 \delta(\omega) \nonumber \\
    &\quad+ T \eta \omega_s \int_{-\infty}^\infty \dd \omega'  \hat{a}_s^\dagger(\omega') \hat{a}_s(\omega - \omega') \nonumber \\
    &\quad+ \sqrt{2 \pi T(1-T) \eta \omega_\mathrm{LO} \omega_s} \qty( \alpha_\mathrm{LO}^\dagger \hat{a}_s(\omega + \Delta_\mathrm{LO}) + \alpha_\mathrm{LO} \hat{a}_s^\dagger(\omega - \Delta_\mathrm{LO})) \biggr]. \label{eqSIderiv:detector_voltage}
\end{align}
We see that the first term peaks at zero frequency and does not contain the signal, so we drop it for the rest of the derivation. $\hat{a}_s(\omega)$ peaks at $\mp \omega = \Delta_{21} \approx \Omega_m^{r/b}$ (see \cref{eqSIderiv:S23}), such that the second term is far outside the detector bandwidth ($\SI{400}{\mega\hertz}$) and we can also neglect it. The third term, however, is shifted into the detector bandwidth by mixing with the local oscillator, so we keep it.
Before we consider the detection of the generated voltage by the electrical spectrum analyzer, we now include the effect of the balanced detection scheme. The amplitude of the mode impinging onto the second detector is given by
\begin{align}
    \hat{a}_{det2} = \sqrt{T} \hat{a}_\mathrm{LO} - \sqrt{1-T}\sqrt{\eta}\hat{a}_{2/1,out}^\mathrm{ext2},
\end{align}
where we note the swapped amplitude transmission between both signals, and the negative sign of the signal term compared to \cref{eqSIderiv:detector_mode}. This comes from the phase shift that one input experiences upon reflection from a dielectric beamsplitter. The relevant term for the detector voltage $V_2$ will then look identical to the last term in  \cref{eqSIderiv:detector_voltage}, except for a global minus sign. In a balanced detector, the photocurrents generated from both photodiodes are subtracted from each other, and the difference current is subsequently converted to a voltage using a transimpedance amplifier to produce a final voltage output of 
\begin{align}
    V(\omega) &= V_1(\omega) - V_2(\omega) = 2 V_1(\omega) \nonumber \\
    &= 2 G \hbar\sqrt{2 \pi T(1-T) \eta \omega_\mathrm{LO} \omega_s} \qty( \alpha_\mathrm{LO}^\dagger \hat{a}_s(\omega + \Delta_\mathrm{LO}) + \alpha_\mathrm{LO} \hat{a}_s^\dagger(\omega - \Delta_\mathrm{LO})) \biggr].
\end{align}

\subsubsection{Signal as displayed on the electrical spectrum analyzer}

The power displayed on an electrical spectrum analyzer is proportional to the symmetrized power spectral density of its input voltage $\Bar{S}_{VV}(\omega) = \frac{1}{2}\qty(S_{VV}(\omega) + S_{VV}(-\omega))$. The unsymmetrized power spectral density of the voltage output from the balanced detector is given by
\begin{align}
    S_{VV}(\omega) &= \int_{-\infty}^\infty \dd \omega' \left\langle V^\dagger(\omega) V(\omega') \right\rangle \\
    &= \underbrace{4 G^2 T(1-T) \eta 2\pi \hbar^2 \omega_\mathrm{LO} \omega_s \abs{\alpha_\mathrm{LO}}^2}_{\equiv \widetilde{\beta}} \int_{-\infty}^\infty \dd \omega' \qty[
    \left\langle \hat{a}_s^\dagger(\omega - \Delta_\mathrm{LO}) \hat{a}_s(\omega' + \Delta_\mathrm{LO}) \right\rangle +
    \left\langle \hat{a}_s(\omega + \Delta_\mathrm{LO}) \hat{a}_s^\dagger(\omega' - \Delta_\mathrm{LO}) \right\rangle]
\end{align}
where from the first to the second line we dropped two of the four total cross-terms, since the correlator of any bath operator (or its Hermitian conjugate) with itself is zero. Also, we assume that all optical and mechanical baths are uncorrelated to each other. Going forward, we use the same arguments, as well as the correlators corresponding to negligible optical bath occupancy and mechanical mode occupancy $n_{th}$:
\begin{align}
    \left\langle \hat{a}_{in,k}^\dagger(\omega) \hat{a}_{in,k}(\omega') \right\rangle &= 0
    & \left\langle \hat{b}_\mathrm{in}^\dagger(\omega) \hat{b}_\mathrm{in}(\omega') \right\rangle &= n_{th} \delta(\omega + \omega') \nonumber \\
    \left\langle \hat{a}_{in,k}(\omega) \hat{a}_{in,k}^\dagger(\omega') \right\rangle &= \delta(\omega + \omega')
    & \left\langle \hat{b}_\mathrm{in}(\omega) \hat{b}_\mathrm{in}^\dagger(\omega') \right\rangle &= (n_{th} + 1) \delta(\omega + \omega')
\end{align}
where $k$ goes over the different internal and external optical baths. We also note that $\left\langle \hat{A}^\dagger(\omega - \Delta_\mathrm{LO}) \hat{A}(\omega' + \Delta_\mathrm{LO})) \right\rangle = \left\langle \hat{A}^\dagger(\omega) \hat{A}(\omega') \right\rangle$, and
\begin{align}
    \mathcal{S}_{ij}^\dagger(\omega' - \Delta_\mathrm{LO}) ~ \mathcal{S}_{ij}(\omega + \Delta_\mathrm{LO}) ~ \delta(\omega + \omega')
    &= \qty(\mathcal{S}_{ij}(\omega' + \Delta_\mathrm{LO}))^\dagger ~ \mathcal{S}_{ij}(\omega + \Delta_\mathrm{LO}) ~ \delta(\omega + \omega') \nonumber \\
    &= \abs{\mathcal{S}_{ij}(\omega + \Delta_\mathrm{LO})}^2.
\end{align}
Inserting what we found previously for $\hat{a}_s(\omega)$ thus gives
\begin{align}
    S_{VV}^{r/b}(\omega) &= \widetilde{\beta} \qty[ \abs{\mathcal{S}_{23}^{r/b}(\pm\omega + \Delta_\mathrm{LO}^{r/b})}^2 (n_{th} + 1) + \abs{\mathcal{S}_{23}^{r/b}(\mp\omega + \Delta_\mathrm{LO}^{r/b})}^2 n_{th} + \sum_{j=1,2,4}\abs{\mathcal{S}_{2j}^{r/b}(\omega + \Delta_\mathrm{LO}^{r/b})}^2].
\end{align}
As shown in \cref{SI:Thermometry_cavity_signal}, the following two relations for the scattering matrix elements hold:
\begin{align}
    \sum_{j=1,2,3,4}\abs{\mathcal{S}_{2j}^r(\omega + \Delta_\mathrm{LO}^r)}^2 &= 1 \\
    - \abs{\mathcal{S}_{23}^b(\omega + \Delta_\mathrm{LO}^b)}^2 + \sum_{j=1,2,4}\abs{\mathcal{S}_{2j}^b(\omega + \Delta_\mathrm{LO}^b)}^2 &= 1
\end{align}
This leaves us with the final expression for the power spectral density of the voltage generated by the balanced detector
\begin{align}
\label{eqSIderiv:PSD}
    S_{VV}^{r/b}(\omega) = \widetilde{\beta} \qty[1 + N_{r/b} \qty( \abs{\mathcal{S}_{23}^{r/b}(\omega + \Delta_\mathrm{LO}^{r/b})}^2+ \abs{\mathcal{S}_{23}^{r/b}(-\omega + \Delta_\mathrm{LO}^{r/b})}^2) ],
\end{align}
where $N_r = n_{th}$ ($N_b = n_{th}+1$) for red (blue) pumping captures the well-known asymmetry of the two scattering signals. The constant offset contributes to the shot noise level. For an accurate prediciton of the true shot noise level, one would have to treat the losses leading to the collection efficiency $\eta$ as an effective beamsplitter onto which vacuum noise is impinging from the other port, but we don't do this here. We note that $S_{VV}^{r/b}(\omega)$ is already symmetric in frequency, such that $\Bar{S}_{VV}^{r/b}(\omega) = S_{VV}^{r/b}(\omega)$. Looking back at \cref{eqSIderiv:S23_mag2}, we observe that $\abs{\mathcal{S}_{23}^{r/b}(\omega)}^2$ peaks at $\omega = \pm \Delta_{21} \approx \pm \Omega_m^{r/b}$. Thus, for red (blue) pumping, the $\abs{\mathcal{S}_{23}^{r/b}(\pm\omega + \Delta_\mathrm{LO}^{r/b})}^2$ term in \cref{eqSIderiv:PSD} peaks at positive $\omega$, the other one will not be shown in the spectrum. We also recall that in the blue pumping case, the sign of the frequency argument in $\Omega_m^b$ and $\Gamma_m^b$ is flipped (cf. \cref{eqSIderiv:Omega_mb_signflip,eqSIderiv:Gamma_mb_signflip}). So $\Omega_m^{r/b}$ and $\Gamma_m^{r/b}$ are centered at $\pm\Delta_{21}$, the same frequency as where $\abs{\mathcal{S}_{23}^{r/b}(\omega)}^2$ peaks, as expected.

The power displayed by the spectrum analyzer is $\Bar{S}_{VV}(\omega) / R_L$ where $R_L = \SI{50}{\ohm}$ is the load resistance, integrated over the bandwidth of the intermediate filter of the spectrum analyzer, i.e. the resolution bandwidth $RBW$. We operate in the limit where the resolution bandwidth is much smaller than the width of the signal peak ($\SI{5}{\kilo\hertz}$ vs. $\sim\SI{50}{\kilo\hertz}$), such that we can replace the integration by simply multiplying with the resolution bandwidth. The final expression for the power displayed on the spectrum analyzer is thus:
\begin{align}
    \label{eqSIderiv:P_ESA}
    P_\mathrm{ESA}^{r/b} (\omega) = \underbrace{2\pi \frac{\mathrm{RBW}}{R_L} 4 G^2 \eta T (1-T)}_{\equiv \beta} \hbar \omega_{2/1} P_\mathrm{LO} \qty[ 1 + g_{r/b}^2
    \frac{\kappa_{2/1}^\mathrm{ext2}}{\qty(\frac{\kappa_{2/1}}{2})^2 + \qty(\omega - (\Delta_{21} - \Delta_\mathrm{LO}) )^2}
    \frac{\Gamma_m N_{r/b}}{\qty(\frac{\Gamma_m^{r/b}}{2})^2 + \qty(\omega - (\Omega_m^{r/b} - \Delta_\mathrm{LO}) )^2}],
\end{align}
where $P_\mathrm{LO} = \hbar \omega_\mathrm{LO} \abs{\alpha_\mathrm{LO}}^2$ and $\Delta_\mathrm{LO} \equiv \abs{\Delta_\mathrm{LO}^{r/b}}$. We observe that the scattering peak is located at $\Delta_{21}-\Delta_\mathrm{LO}$ for both the Stokes- and Anti-Stokes scattering process. Note that for the convention of the Fourier transform used here (unitary form, angular frequency), Parseval's theorem takes the form
\begin{align}
    \int_{-\infty}^\infty \dd t \abs{x(t)}^2 = \frac{1}{2\pi} \int_{-\infty}^\infty \dd \omega \abs{X(\omega)}^2,
\end{align}
such that the total power in the spectrum when integrating over \emph{ordinary} frequency is a factor $2\pi$ smaller.

\subsection{Corrections to thermometry signals}
\label{SI:Thermometry_corrections}
Here we discuss the procedure for obtaining the thermal mode occupations $n_{th}$ from the measured heterodyne spectra, as shown in \cref{fig:3_Thermometry,fig:4_LaserFridgeHeating} of the main paper. In principle, the power of sideband asymmetry thermometry is that it allows us to extract $n_{th}$ by taking the ratio of second term in \cref{eqSIderiv:P_ESA} for the red and blue pumping cases, removing the need to calibrate the prefactors. In practice, however, one has to carefully correct for any differences in the measurement conditions for the two cases. To a large extent, these corrections are similar to what is done in earlier works on optomechanical sideband asymmetry thermometry \cite{Delic2019,Weinstein2014a,Peterson2016}, but the fact that we use two optical modes introduces some differences, which we will describe here.

\subsubsection{Red- and blue-pumped integrated signals}

As derived in \cref{SI:Thermometry_theory}, the observed power on the electrical spectrum analyzer is composed of a shot noise term and a term due to the peak from (anti-)Stokes scattering. By using a strong local oscillator, we work in the regime where the noise floor is dominated by shot noise, but a slight level of technical noise still contributes. This noise floor is not completely flat, so we fit the noise floor with a second order polynomial and subtract it from the total signal.

The power in the signal peak is then
\begin{align}
\label{eqSIcorr:signal_in}
    P_\mathrm{peak}^{r/b} (\omega) = \beta \hbar \omega_{2/1} P_\mathrm{LO} g_{r/b}^2
    \frac{\kappa_{2/1}^\mathrm{ext2}}{\qty(\frac{\kappa_{2/1}}{2})^2 + \qty(\omega - (\Delta_{21} - \Delta_\mathrm{LO}) )^2}
    \frac{\Gamma_m N_{r/b}}{\qty(\frac{\Gamma_m^{r/b}}{2})^2 + \qty(\omega - (\Omega_m^{r/b} - \Delta_\mathrm{LO}) )^2},
\end{align}
where $\beta$ contains all constants that do not vary between red and blue pumping measurements.

We work in the regime where the optical linewidth is much larger than the mechanical linewidth ($\sim\SI{2}{\mega\hertz} \gg \sim \SI{50}{\kilo\hertz}$. This allows us to integrate \cref{eqSIcorr:signal_in} over a frequency range $\pm \delta$ around the peak position $\Omega_m^{r/b} - \Delta_\mathrm{LO}$ that covers the mechanical linewidth, but is still small with respect to the optical linewidth, i.e. $\Gamma_m^{r/b} \ll \delta \ll \kappa_{2/1}$. This way, we can neglect the frequency dependence of the first term and approximate
\begin{align}
    I^{r/b} &\equiv \int_{\Omega_m^{r/b} - \Delta_\mathrm{LO} - \delta}^{\Omega_m^{r/b} - \Delta_\mathrm{LO} + \delta} \dd \omega' P_\mathrm{peak}^{r/b} (\omega') \nonumber \\
    &\approx \beta \hbar \omega_{2/1} P_\mathrm{LO} g_{r/b}^2
    \frac{\kappa_{2/1}^\mathrm{ext2}}{\qty(\frac{\kappa_{2/1}}{2})^2 + \qty(\Omega_m^{r/b} - \Delta_{21} )^2} \frac{\Gamma_m}{\Gamma_m^{r/b}} N_{r/b}.
\end{align}
 We can rewrite $g_{r/b}^2 = g_0^2 \frac{4\kappa_{1/2}^\mathrm{ext1}}{\kappa_{1/2}^2} \abs{\alpha_{p,in}}^2$, where we assumed the pump tone to be on resonance, which is ensured by the PDH lock. Thus,
\begin{align}
    I^{r/b} = \Bar{\beta} P_p P_\mathrm{LO} \frac{1}{\qty(\frac{\kappa_{2/1}}{2})^2 + \qty(\Omega_m^{r/b} - \Delta_{21} )^2} \frac{\kappa_{1/2}^\mathrm{ext1}\kappa_{2/1}^\mathrm{ext2}}{\Gamma_m^{r/b} \kappa_{1/2}^2} N_{r/b}
    \label{SI:Corr:eq:IntegralRB}
\end{align}
where $P_p \approx \hbar \omega_{2/1} \abs{\alpha_{p,in}}^2$ is the pump power, and $\Bar{\beta} = 4 \beta g_0^2 \Gamma_m$. All variables in this expression, other than $N_{r/b}$, can be characterized through separate measurements: $P_p$ and $P_\mathrm{LO}$ are monitored on photodetectors, $\kappa_{1/2}^\mathrm{ext1}, \kappa_{1/2}^\mathrm{ext2}$ cannot be individually determined, but the ratios of these coupling rates between one mode and the next can be obtained from fits to the cavity reflection spectra (see \cref{SI:Optical_input_coupling_characterization}), and the other parameters are obtained from fits to OMIT and OMIA spectra.
Thus, we can divide these parameters out and obtained the corrected integrated signals
\begin{align}
    I_\mathrm{corr}^r &= \Bar{\beta} n_{th} \label{eqSIcorr:Icorr_r} \\
    I_\mathrm{corr}^b &= \Bar{\beta} (n_{th} + 1). \label{eqSIcorr:Icorr_b}
\end{align}
We solve this for the thermal mode occupation $n_{th}$ and obtain
\begin{align}
\label{eqSIcorr:nb}
    n_{th} = \frac{1}{\frac{I_\mathrm{corr}^b}{I_\mathrm{corr}^r} - 1}.
\end{align}
This assumes that $n_{th}$ is the same for both red and blue pumping measurements, which is ensured by waiting for the fridge to return to similar temperatures and pressures after the pulse tube is turned off for a previous measurement. For measurements during fridge warmup, this assumption is not true anymore, which is why we employ a different method to extract $n_{th}$, see \cref{SI:Warmup_measurements}.

Finally, we note that when we show the corrected thermometry signals in \cref{fig:3_Thermometry}a-f, instead of dividing out all the prefactors mentioned above, we multiply the red spectrum (not including its baseline of 1) by the \emph{ratio} of the prefactors for the blue and red data. This ensures that the amplitude of both peaks can still be compared to the baseline of 1 to estimate our signal-to-noise ratio, while the blue/red asymmetry is given by the ratio of the areas under the blue and red peaks, respectively.

\subsubsection{Optical input coupling characterization}
\label{SI:Optical_input_coupling_characterization}
According to \cref{SI:Corr:eq:IntegralRB}, we need to correct for differences in the external coupling rates $\kappa_{1/2}^\mathrm{ext1}, \kappa_{1/2}^\mathrm{ext2}$. As the integrated signal for red (blue) pumping $I^r$ ($I^b)$ is proportional to $\kappa_{1}^\mathrm{ext1} \kappa_{2}^\mathrm{ext2}$ ($\kappa_{2}^\mathrm{ext1} \kappa_{1}^\mathrm{ext2}$), the ratio $I^b / I^r$ must be multiplied with a factor $\kappa_{1}^\mathrm{ext1} \kappa_{2}^\mathrm{ext2} / \kappa_{2}^\mathrm{ext1} \kappa_{1}^\mathrm{ext2}$ to obtain $I^b_{corr} / I^r_{corr}$. 
While, in principle, the external coupling rates can be determined from the cavity reflection and transmission spectra, as done in \cite{Kharel2022}, this is not possible in the presence of unknown losses due to misalignments or e.g. fiber transmission. We can, however, determine the \emph{ratio} of external coupling rates of our two modes, which is sufficient for the purpose of this correction. 

As discussed in \cref{SI:Alignment_procedure}, in the presence of fiber losses and a mismatch between input or output optics and the cavity modes, the cavity reflection is described by \cref{SI:eq:R1}. This can be rewritten as 
\begin{align}
    R_1(\Delta) &= \left| s_{12,1} s_{21,1} + s_{11,1} \right|^2 \left| 1-\frac{s_{12,1} s_{21,1}}{s_{12,1} s_{21,1} + s_{11,1}} \frac{\kappa^\mathrm{ext1}}{\kappa/2 -i\Delta} \right|^2 \\
    &=  R_{1,\Delta\gg\kappa} \left| 1- S'e^{-i\phi} \frac{\kappa^\mathrm{ext1}}{\kappa/2 -i\Delta} \right|^2, \label{SI:eq:R1_fano}
\end{align}
with $R_{1,\Delta\gg\kappa}$ the reflection far from resonance and $S'=|(s_{12,1} s_{21,1})/(s_{12,1} s_{21,1} + s_{11,1})|$. \Cref{SI:eq:R1_fano} describes a Fano resonance \cite{Novotny2012}, with $\phi$ the Fano phase that determines the asymmetry of the resonant feature.
By fitting this expression to the reflection spectra of our red and blue cavity modes, we can obtain $R_{1,\Delta\gg\kappa}$, $\phi$, $\kappa$, $\omega_0$ and $S'\kappa^\mathrm{ext1}$. Thus we see that we cannot uniquely determine the input coupling rate through a reflection fit. If we assume, however, that $S_1$ and $S_2$ are frequency-independent within the frequency range spanning our two cavity modes, we can determine the ratio $\kappa_1^\mathrm{ext1}/\kappa_2^\mathrm{ext1}$ through a fit of both cavity modes. Similarly, by fitting the reflection spectra of both modes with the laser entering from port 2 (the `back' side of the cavity), we can find the ratio $\kappa_1^\mathrm{ext2}/\kappa_2^\mathrm{ext2}$. We therefore record the reflection spectra of both our red and blue modes, illuminated through port 1 and port 2 to obtain these ratios necessary for the thermometry signal correction. Such spectra are recorded for every cavity setting at which we do thermometry, i.e. after alignment and tuning of the mode pair to the displacement-insensitive point (see \cref{SI:Displacement-insensitive_point}), both at 4K and at mK temperatures.


\subsection{Warmup measurements}
\label{SI:Warmup_measurements}

For the main results of the paper, the phonon mode occupation is extracted by observing the asymmetry between the (corrected) integrated Stokes and anti-Stokes scattering signals (see \cref{eqSIcorr:nb}). This assumes, however, that the phonon mode temperature is the same between both measurements, which is not true during fridge warmup. Thus, we employ a different method which is based on interpolating the signals from a reference measurement pair before warmup.

We take a pair of measurements (red and blue pumping) shortly before starting the fridge warmup, and extract the thermal mode occupation $n_{th}$ in the usual way according to \cref{eqSIcorr:nb}. We call this the reference measurement pair. For subsequent measurements at higher temperatures, we expect the corrected integrated signals to then scale with $n_{th}$ for red pumping, $n_{th}+1$ for blue pumping, respectively (see \cref{eqSIcorr:Icorr_r,eqSIcorr:Icorr_b}). Thus, by assuming $\Bar{\beta}$ stays constant during fridge warmup, we can extract $n_{th}$ from just a single measurement via
\begin{align}
\label{eqSIwarmup:interpolation}
    n_{th} =
    \begin{cases}
        n_{th}^\mathrm{ref} \frac{I_\mathrm{corr}^r}{I_\mathrm{corr,ref}^r} & \text{for red pumping} \\
        (n_{th}^\mathrm{ref}+1) \frac{I_\mathrm{corr}^b}{I_\mathrm{corr,ref}^b} - 1 & \text{for blue pumping.}
    \end{cases}
\end{align}

\begin{figure}
    \centering
    \includegraphics{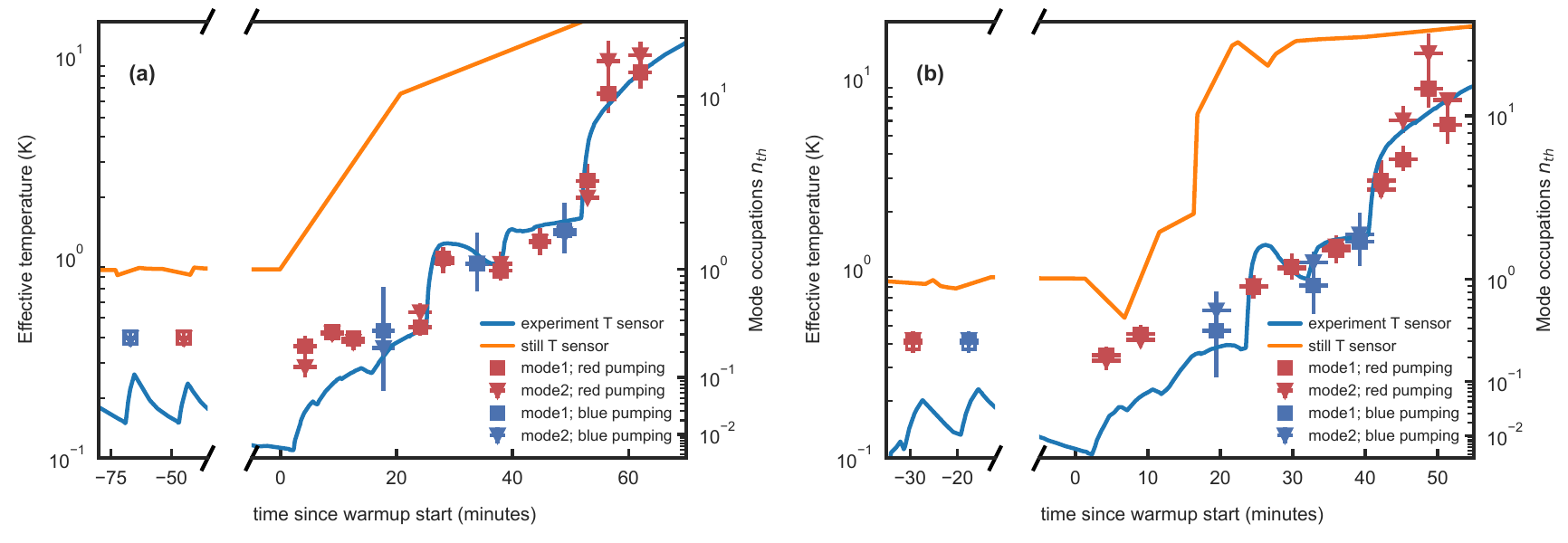}
    \caption{\textbf{Warmup analysis with one reference measurement pair.} Instead of 3 reference measurement pairs, just the one before the start of the warmup is defined, as indicated by the hollow markers.
    \textbf{(a)}~Data is identical to that presented in the main text.
    \textbf{(b)}~Data shown is from another cooldown as the data in the rest of the paper, during which the experiment was surrounded by eccosorb foam.
    }   \label{fig:SI_warmup_1refpairs}
\end{figure}

During fridge warmup, there are period of time where the experiment temperature stays on a plateau that is long enough for two measurements, such that we can declare them as a new reference measurement pair (see hollow markers in \cref{fig:4_LaserFridgeHeating}c. The measurements coming afterwards then refer back to this reference pair for the interpolation according to \cref{eqSIwarmup:interpolation}. To test whether defining these new reference measurement pairs is valid, we perform the same analysis on the same data but just using the reference measurements from before the start of the fridge warmup. We observe the same qualitative behaviour as explained in the main text, see \cref{fig:SI_warmup_1refpairs}a. In a separate cooldown than the one in which all data in the main text was taken, eccosorb foam was added into the copper heat shield that surrounds the experiment. This was done in the hope that this would better thermalize the blackbody environment to the experiment temperature. However, no significant effect on the crystal temperatures was observed, as can be observed from the qualitatively same results in \cref{fig:SI_warmup_1refpairs}b.

\subsection{An equivalent thermal circuit model to describe the crystal temperature}
\label{SI:ThermalModel}
\begin{figure}
    \centering
    \includegraphics{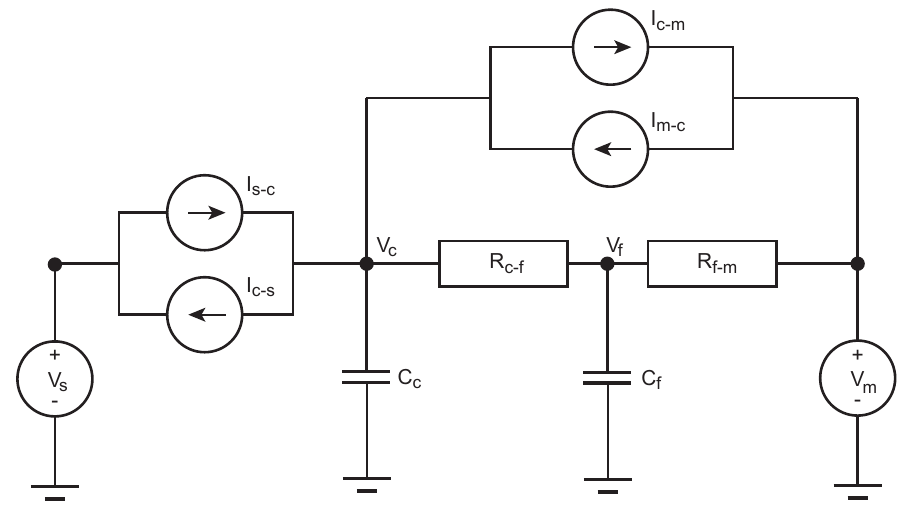}
    \caption{\textbf{Equivalent thermal circuit describing heat exchange between the crystal, its mount and the still plate.} This circuit models conductive heat exchange between crystal and its mount, through the front mirror, as well as blackbody radiation exchange between crystal and mount, and between crystal and still plate. 
    Temperatures correspond to voltages, heat flows to currents, thermal resistances to resistors and heat capacitances to capacitors. 
    }   \label{SI:fig:ThermalCircuit}
\end{figure} 

To test whether blackbody radiation by a higher-temperature stage could explain the elevated occupations we observe at mK temperatures in \cref{fig:3_Thermometry} of the main paper, we develop an equivalent thermal circuit model and use it to fit the data from our warmup measurement, shown in \cref{fig:4_LaserFridgeHeating}c of the main paper.
We assume that such blackbody radiation will be dominated by the still plate, which is the warmest plate that can easily exchange radiation with our experiment, since radiation from higher plates is shielded by the still plate and still can. 
Heat exchange between crystal and its mount (the temperature of which we monitor with our experiment thermometer) happens either through blackbody radiation or through conductive exchange through the front mirror and teflon spacer that the crystal is clamped to. 

Under these assumptions the equivalent thermal circuit for our system can be drawn as in \cref{SI:fig:ThermalCircuit}. Temperatures of the crystal, front mirror (the facet that is facing the crystal), mount, and still plate are represented by voltages $V_c, V_f, V_m$ and $V_s$, respectively. The thermal resistance between front mirror and mount (front mirror and crystal) is given by the resistance $R_{f-m}$ ($R_{c-f}$), while front mirror and crystal each have a thermal capacitance (i.e. heat capacity) of $C_f$ and $C_c$. 
The voltages $V_m$ and $V_s$ are set by voltage sources to the values we measure on our experiment and still thermometers, respectively. 
Blackbody radiation from, for example, still plate to crystal is described by a current source with current 
\begin{equation}
    I_{s-c}=b_{s-c} V_s^4, \label{SI:eq:Isc}
\end{equation}
in accordance with the Stefan-Boltzmann law, with $b_{s-c}=\sigma F_{s-c} \epsilon_s A_s$. Here, $\sigma$ is the Stefan-Boltzmann constant, $\epsilon_s$ the still plate emissivity (between 0 and 1), $A_s$ the still plate area and $F_{s-c}$ a fraction of the still plate radiation that is absorbed by the crystal. Equivalent relations hold for the blackbody radiation currents $I_{s-c}, I_{c-m}$ and $I_{m-c}$. 

Now we make a simplifying assumption to reduce the number of free parameters.
Although in general $\epsilon$ and $F$ depend weakly on temperature, as the blackbody emission spectrum shifts with temperature and the absorptivity and emissivity of any object depend on wavelength, we will ignore this dependence here and assume them to be constant. 
Furthermore, the second law of thermodynamics dictates that if two objects are the same temperature, no net heat exchange between them can take place, i.e. $I_{s-c}(V_s)= I_{c-s}(V_c)$ for $V_s=V_c$ and $I_{m-c}(V_m)= I_{c-m}(V_c)$ for $V_m=V_c$. Therefore, under the assumption that  $\epsilon$ and $F$ are temperature-independent, we find that $b_{c-s}=b_{s-c}$ and $b_{c-m}=b_{m-c}$. 

We also consider that the thermal resistances $R_{f-m}, R_{c-f}$ are temperature-dependent. 
We take $R_{f-m}$ to be given by the sum of the contact resistances $R_{f-m,0}$ between mount and front mirror (assumed temperature-independent) and the thermal resistance $R_{f-m,f}$ of the fused silica front mirror, which scales with temperature as $R_{f-m,f}=R_{f-m,1}/V_f$ up to \SI{3}{\kelvin} \cite{Damon1973} (ignoring temperature gradients in the mirror), with $R_{f-m,1}$ a constant.
We similarly take $R_{c_f}$ to be given by the sum of the contact resistances $R{c-f,0}$ between the crystal, its teflon spacer and the front mirror (again assumed temperature-independent) and the thermal resistance $R_{c-f,t}$ of the \SI{0.2}{\milli\meter}-thick
teflon spacer, which scales with temperature as $R_{c-f,t}=R_{c-f,2}/V_t^2$ up to $\sim\SI{4}{\kelvin}$ \cite{Scott1972}, with $R_{c-f,2}$ a constant and $V_t$ the voltage (temperature) of the teflon. 
Since the teflon or front mirror temperatures are not known, we simplify our analysis by assuming that these thermal resistances depend on $V_m$ instead, i.e. $R_{f-m,f}=R_{f-m,1}/V_m$ and $R_{c-f,t}=R_{c-f,2}/V_m^2$. This corresponds to a kind of 'worst-case' estimate for the resistances $R_{f-m}, R_{c-f}$, since at low temperatures $V_m$ is lower than $V_c$ (and so also lower than $V_t$ or $V_f$), leading to an overestimation by the model of the thermal resistance there. Overestimating this resistance will imply that at low temperature, the model predicts crystal temperatures to lie closer to the still temperatures than they should. We will see that this approximation will therefore not change the conclusions we draw from this analysis. 

Having set up the equivalent thermal circuit, we apply the Kirchoff current law to the node at voltage $V_c$ and that at $V_f$ to find
\begin{align}
    C_c \dot{V}_c &= (I_{s-c}-I_{c-s}) + (I_{m-c}-I_{c-m}) + \frac{V_f-V_c}{R_{c-f}} \label{SI:eq:Vc_dot}\\
    C_f \dot{V}_f &= \frac{V_c-V_f}{R_{c-f}} + \frac{V_m-V_f}{R_{f-m}}. \label{SI:eq:Vf_dot}
\end{align}
In general these coupled non-linear first-order differential equations could be solved numerically by, for example, the Runge-Kutta method. Here we simplify our analysis further by assuming that the dynamics are faster than our measurement time such that we measure our crystal always in the steady state, i.e. $\dot{V}_c=\dot{V}_f=0$. This seems reasonable, given how fast the crystal temperature follows the sharp increase in mount temperature at $t\sim \SI{55}{\minute}$ in \cref{fig:4_LaserFridgeHeating}c of the main paper.
We may then solve \cref{SI:eq:Vf_dot} for $V_f$ and plug that into \cref{SI:eq:Vc_dot} to eliminate $V_f$ (which is unknown) and obtain
\begin{equation}
    V_c= V_m + R_{c-m} \left( (I_{s-c}-I_{c-s}) + (I_{m-c}-I_{c-m}) \right),
\end{equation}
where $R_{c-m}=R_{c-f}+R_{f-m}$. If we now plug in the temperature dependencies of the blackbody currents as in \cref{SI:eq:Isc} (and similarly for the other currents), we obtain the following quartic equation for $V_c$:
\begin{equation}
    R_{c-m}(b_{s-c}+b_{m-c})V_c^4 + V_c - \left(V_m + R_{c-m} \left( b_{s-c}V_s^4+b_{m-c}V_m^4 \right) \right) = 0. \label{SI:eq:Vc_quartic}
\end{equation}
Here, as discussed above, the thermal resistance $R_{c-m}$ is given as the sum of three terms with different temperature dependencies, i.e.
\begin{equation}
    R_{c-m}= R_{c-m,0} + \frac{R_{c-m,1}}{V_m} +\frac{R_{c-m,2}}{V_m^2},
\end{equation}
with $R_{c-m,0},R_{c-m,1}$ and $R_{c-m,2}$ constants. 

To test whether this model can predict the temperatures we measure for our mechanical modes (taken to represent the crystal temperature) during the fridge warmup, we try to fit the model to the temperature data shown in \cref{fig:4_LaserFridgeHeating}c of the main paper. At each time when we have a temperature measurement of the crystal, we find the corresponding mount and still temperatures by interpolating the temperature sensor data. This gives us a data set of input voltages (temperatures) $\{V_m,V_s\}$ and resulting crystal voltages $\{V_c\}$. We find crystal voltages from our model for each combination of $V_m$ and $V_s$ by finding the four roots of \cref{SI:eq:Vc_quartic}, postselecting the positive real roots between $V_m$ and $V_s$ (values outside that range are unphysical) and then picking the smallest root if there are multiple candidates. We also tried picking the largest root but it doesn't change the result.
Initial guesses for the fit parameters $R_{c-m,0}, R_{c-m,1}, R_{c-m,2}, b_{s-c}$ and $b_{m-c}$ are based on literature values of low-temperature conductivities of Teflon and fused silica, and the physical dimensions of the crystal and other elements involved. 

An automated minimization routine did not manage to find a good fit to our data. We therefore manually adjusted the fit parameters to see if we could make the model fit the data. We were able to find parameters for which we were in the regime dominated by conductive heat transport, where the crystal always follows the mount temperature. By increasing $b_{s-c}$ we can go to a regime where radiation from the still plate dominates and the crystal follows the still temperature closely. Picking an intermediate value for $b_{s-c}$ brings us into the regime where the crystal temperatures lie somewhere in between mount and crystal. However, regardless what values we choose for our parameters in that region, we cannot reproduce the measured temperature behaviour of our crystal, which is to follow the mount temperatures closely at high mount temperatures ($T_m>\sim\SI{400}{\milli\kelvin}$), yet to stabilize around a crystal temperature of $T_c=\sim\SI{400}{\milli\kelvin}$ when the mount temperature drops below $\sim\SI{400}{\milli\kelvin}$. We also tried setting $R_{c-m}$ to infinity and increasing $b_{m-c}$ to balance black body radiation from the still and to the mount, but also this cannot explain our data. 

We believe that the reason that our model cannot predict the qualitative temperature behaviour we measure, is the following: let us assume that black body radiation from the still causes the elevated crystal temperatures at low mount temperatures. That means that at those low temperatures, heating by this blackbody radiation and cooling by the mount balance each other. However, blackbody radiation grows proportional to $T_s^4$, while the heat flow through $R_{c-m}$ can grow at most proportional to $T_m^2$ (in the case where $R_{c-m}$ is dominated by $R_{c-m,2}$). Thus, as time progresses during the warmup and both still and mount rise in temperature, heat flow into the crystal by blackbody radiation from the still must grow compared to the removal of heat from the crystal through $R_{c-m}$, thus bringing the crystal temperature closer to that of the still. 
Even if the heat exchange between mount and crystal were dominated by blackbody radiation (and would thus be proportional to $T_m^4$), we still cannot obtain the desired behaviour. In that case, neglecting the contributions of $I_{m-c}$ and $I_{c-s}$, the ratio $T_s^4/T_m^4$ determines the ratio of input to output heat flows for the crystal. So if $T_s$ and $T_m$ grow by the same factor, this ratio is conserved. However, we see in \cref{fig:4_LaserFridgeHeating}c that in the beginning of the warmup (around $t\sim10-20$ min), the still stage grows more rapidly in temperature, while only in the last phase the mount temperature grows faster. Such a pure blackbody radiation model thus always predicts the crystal temperature to approach the still temperature more closely in this early phase of the warmup, which is not what we observe. 

We thus conclude that radiation by an object that is at the temperature of the still plate cannot explain why our crystal temperatures are larger than those of the mount. Neither can radiation by higher-temperature stages, since those warm up even more rapidly than the still stage during the warmup. 
A source of heat that would be consistent with our data, however, is radiation by a poorly thermalized still shield. This shield surrounds our experiment, it is long and its walls are thin, and it is being heated by radiation from the $\SI{4}{\kelvin}$ stage and its shield, so it is likely that it is indeed not always at the same temperature as the still stage. One would then expect it to be at higher temperatures than the still plate during normal operation, but lagging behind the increasing temperature of the still plate during the fridge warmup, which could help explain the behaviour we see. 

\subsection{Laser phase noise measurements}
\label{SI:PhaseNoise}

\begin{figure}[b]
    \centering
    \includegraphics{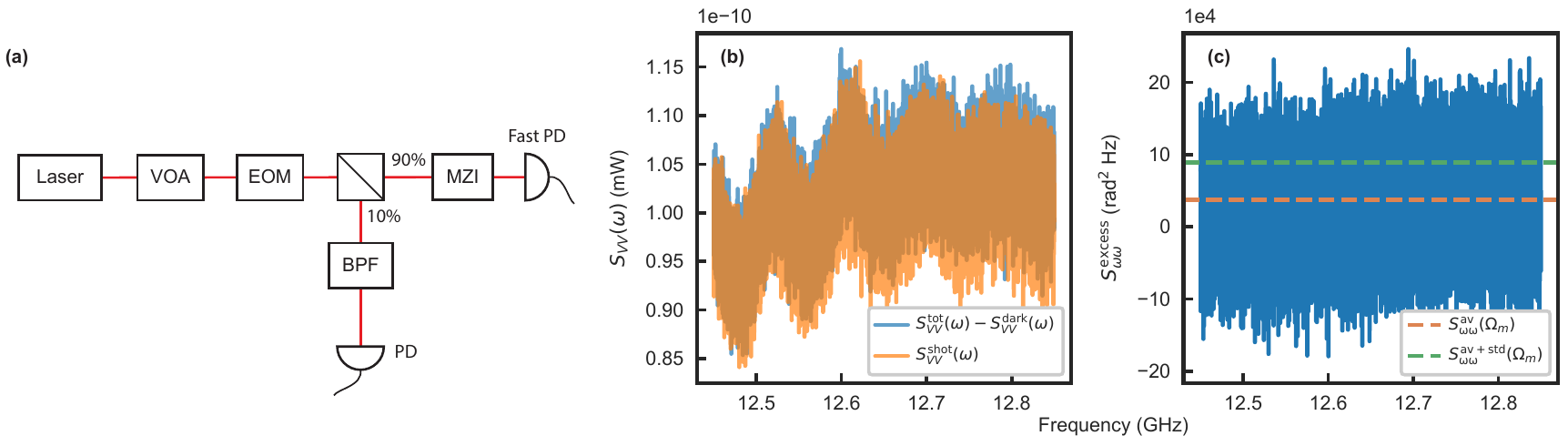}
    \caption{\textbf{Laser frequency noise measurement}. 
    \textbf{(a)}~Setup used to measure contributions of frequency noise in the spectrum. VOA: variable optical attenuator, EOM: electro-optical modulator, MZI: fiber Mach-Zehnder interferometer, PD: photodetector. 
    \textbf{(b)}~Comparing ESA spectra measured with (blue) and without (orange) the MZI, with the same optical power and the detector noise ($\sim 1.7$ dB below signals) subtracted.  
    \textbf{(c)}~Laser frequency noise spectral density, obtained by subtracting the blue and orange traces in (b) and multiplying by $A(\omega)$. Orange and green dashed lines show the average and the average plus one standard deviation, respectively, of the data within a 10 MHz span around $\Omega_m$.
    }   
    \label{SI:fig:Phasenoise}
\end{figure} 

Laser phase noise produces sidebands that can appear in sideband thermometry measurements. If the noise is at the same frequency as the mechanical sideband, it can, depending on relative phase conditions, increase or decrease the measured sideband asymmetry in optomechanical experiments in addition to affecting the actual mechanical mode occupation \cite{Safavi-Naeini2013a}. 


To account for this contribution, we follow the procedure described in \cite{Safavi-Naeini2013a} in order to evaluate the influence of laser phase noise on the detected phonon occupancy in sideband asymmetry experiments. 
We use a calibrated electro-optical phase modulator (EOM) and a Mach-Zehnder interferometer (MZI) to generate a calibrated phase modulation signal that is then used to quantify the phase noise intrinsic to the laser (see \cref{SI:fig:Phasenoise}a). 
The same Toptica CTL 1550 tunable laser at 1550 nm as used for our sideband thermometry measurements is used as the light source in this phase noise measurement, at a power of \SI{673}{\micro\watt}. The EOM is calibrated using a fiber-based narrowband-pass filter. The relative transmission peak size of carrier and sidebands allows us to characterize the modulation depth $\beta$ of our modulator at the desired frequencies around the mechanical mode, using 

\begin{equation}
 \label{eq:SI-bessel}
    \left[\frac{J_1(\beta)}{J_0(\beta)}\right]^2=\frac{P_1}{P_0}.
\end{equation}

\noindent where $P_0$ ($P_1$) is the power in the carrier (first sideband) and $J_k$ are the Bessel functions of the first kind.

The phase modulation created by the EOM leads to frequency noise on the laser, with their noise power spectral densities related by $S_{\omega\omega}(\omega) = \omega^2 S_{\phi\phi}(\omega)$ \cite{Safavi-Naeini2013a}. If we use a sinusoidal modulation of frequency $\omega_\mathrm{mod}$, we create a frequency noise spectrum (in \si{\radian^2 \hertz})

\begin{equation}
    S_{\omega\omega}^\mathrm{EOM}(\omega)=\frac{\pi\omega^2\beta^2}{2}\left[ \delta(\omega-\omega_\mathrm{mod})+\delta(\omega+\omega_\mathrm{mod})\right].
\end{equation}

\noindent The laser light is then passed through a fiber MZI, with the laser locked to its half-max point, which converts the phase noise to amplitude noise. The MZI free spectral range is designed to be 36 GHz, which is well-suited to the noise frequencies of interest, which lie around the mechanical mode frequency of $\approx \SI{12.66}{\giga\hertz}$. 
The MZI transmission is recorded on a fast photodetector (Thorlabs RXM25AF, operated with nominal gain of \SI{1500}{\volt/\watt}), the output voltage of which is recorded on an electronic spectrum analyser (ESA) to produce its power spectral density $S_\mathrm{VV}(\omega)$ that we can write as
\begin{equation}
    S_\mathrm{VV}(\omega) = S_{\omega\omega}(\omega)/ A(\omega). \label{eqSIphasenoise:Aom}
\end{equation}
Here, $A^{-1}(\omega)$ is the conversion factor with which frequency noise is converted to voltage noise by our setup. Since, for the calibration signal, $S_{\omega\omega}(\omega)=S_{\omega\omega}^\mathrm{EOM}(\omega)$ is known, we can determine $A(\omega)$ from the integrated measured power spectral density. 

Having calibrated the setup conversion factor $A(\omega)$, we now proceed to measure the intrinsic laser frequency noise. The laser is sent through the fiber MZI, and we turn off the modulation tone to the EOM. 
The measured ``total noise" spectrum $S^\mathrm{tot}_\mathrm{VV}(\omega)$ on the ESA is shown in \cref{SI:fig:Phasenoise}b and is the sum of the detector dark noise $S^\mathrm{dark}_\mathrm{VV}(\omega)$, laser shot noise $S^\mathrm{shot}_\mathrm{VV}(\omega)$ and the excess (i.e. classical) laser frequency noise $S^{\omega,\mathrm{excess}}_\mathrm{VV}(\omega)$. We confirmed separately that the laser shows no appreciable excess intensity noise near our mechanical frequency, by verifying linear scaling of the intensity noise with power (measured without the MZI). 
We notice the absence of any pronounced peak in the noise spectrum that would indicate relaxation-oscillation induced frequency noise present in other diode lasers \cite{Kippenberg2013,Safavi-Naeini2013a}.

\Cref{eqSIphasenoise:Aom} only applies to the excess frequency noise, not to shot or detector noise. 
We therefore measure the shot noise in a separate measurement where we bypass the MZI and shine the laser onto the same detector, locking it to stabilize it to the same power (within 0.2\%) as used in the phase-sensitive measurement of $S^\mathrm{tot}_\mathrm{VV}(\omega)$. This shot noise (see \cref{SI:fig:Phasenoise}b) is just below the total noise, so we subtract it from the total noise to obtain the excess frequency noise $S^{\omega,\mathrm{excess}}_\mathrm{VV}(\omega)$. 
The conversion coefficient $A(\omega)$ can then be used to convert this excess phase noise to a power spectral density of the frequency noise on the laser, $S_{\omega\omega}^{laser}(\omega)$. The resulting laser frequency noise spectrum found in our experiment is shown in \cref{SI:fig:Phasenoise}c. 
As a conservative estimate of the noise at the mechanical frequency $\Omega_m /2\pi\approx \SI{12.66}{\giga\hertz}$, we take both the average $S^{\mathrm{av}}_\mathrm{\omega\omega}(\Omega_m)$ over a \SI{10}{\mega\hertz} span around $\Omega_m$ as well as this average plus one standard deviation of the noise in that same range, $S^{\mathrm{av + std}}_\mathrm{\omega\omega}(\Omega_m)$. 
This yields $S^{\mathrm{av}}_\mathrm{\omega\omega}(\Omega_m)=\SI{3.7e4}{\radian^2\hertz}$ and $S^{\mathrm{av+std}}_\mathrm{\omega\omega}(\Omega_m)=\SI{8.9e4}{\radian^2\hertz}$.

\subsubsection{Heating of the mechanical resonator}
At the mechanical resonance frequencies, phase noise could contribute to the phonon population as an additional source of noise leading to actual heating in the system. This contribution can be computed using input-output relations and considering the non-zero correlation of the phase-noise \cite{Safavi-Naeini2013a}. In the resolved-sideband limit ($\Omega_m \gg \kappa$) and weak-coupling regime ($\kappa \gg \Gamma_{m,\mathrm{eff}}$), assuming $\Delta_{21}=\Omega_m$ and pumping on the red mode at frequency $\omega_1$, the mode occupation modification induced by the laser phase noise can be calculated as

\begin{equation}
    n_{\phi}^{\mathrm{phonon}}=\frac{\left(\Gamma_{m,\mathrm{eff}}-\Gamma_m \right) }{\Gamma_{m,\mathrm{eff}}} \frac{\kappa^{ext}_2}{\kappa_2}n_{\phi}^{\mathrm{photon}}, \label{eqSIphasenoise:nphonon_by_phasenoise}
\end{equation}

\noindent where $n_\mathrm{\phi}^{\mathrm{photon}}\equiv S_\mathrm{EE}(\Omega_\mathrm{m})$ is number of noise photons present in the light field at a frequency detuned by $\Omega_m$ from the central laser frequency. This can be estimated from the laser frequency noise as
\begin{equation}
    n_{\phi}^{\mathrm{photon}}=\frac{S_\mathrm{\omega\omega}(\Omega_m)}{\Omega_\mathrm{m}^2}|E_0|^2,
\end{equation}  
where $|E_0|$ is the optical input field amplitude in $\sqrt{\mathrm{photons \, s}^{-1}}$. 
Assuming a cooperativity of $C=0.15$ (the maximal value in \cref{fig:3_Thermometry}), $\kappa^{ext}_2 / \kappa_2 \approx 1/2$ and $|E_0|^2\approx 6.4\times 10^{14}$ $\mathrm{photons \, s}^{-1}$ (corresponding to the optical power $P_\mathrm{opt} = \SI{82.5}{\micro\watt}$ used in the thermometry experiment), we find an added phonon number of $n_{\phi}^{\mathrm{phonon}}=\SI{2.5e-4}{}$ or $n_{\phi}^{\mathrm{phonon}}=\SI{5.9e-4}{}$ when using $S^{\mathrm{av}}_\mathrm{\omega\omega}(\Omega_m)$ or $S^{\mathrm{av+std}}_\mathrm{\omega\omega}(\Omega_m)$ for the frequency noise level, respectively. 
We therefore conclude that phonon heating due to laser phase noise is negligible in our thermometry measurements.

\subsubsection{Effect on sideband asymmetry thermometry}
Phase noise can also lead to a change in the mode occupation as inferred from sideband asymmetry thermometry measurements. Depending on the optical input-output relations of the system, either the red-pumped noise spectra experiences noise squashing and the blue detuned data anti-squashing, or vice versa. To estimate the magnitude of this effect in our measurements, we follow the approach in \cite{Safavi-Naeini2013a} to quantify how the inferred occupations differ from the actual occupations. 

As discussed in \cref{SI:Thermometry_corrections}, we infer the mechanical occupation from the ratio of the corrected red- and blue-pumped integrals through \cref{eqSIcorr:nb}, which we can also write as
\begin{align}
    n_{\mathrm{th}}^\mathrm{inf} = \frac{1}{\frac{\Bar{I}^b (1-C)}{\Bar{I}^r(1+C)} - 1} \label{eqSIphasenoise:nth_inf}
\end{align}
where we assumed again that $\Delta_{21}=\Omega_m$ (such that $\Gamma^{r/b}_{m,\mathrm{eff}}=(1\pm C)\Gamma_m$), and where we defined $\Bar{I}^{r/b}$ to represent the integrals of the thermometry signals, corrected for all factors that differ between them except that of the difference in mechanical linewidth, i.e. 
$\Bar{I}^r = \Bar{\beta}  n_{th} / \Gamma^{r}_{m,\mathrm{eff}}$ and 
$\Bar{I}^b = \Bar{\beta} (n_{th} + 1) / \Gamma^{b}_{m,\mathrm{eff}}$.
We note that this description is equivalent to that used in \cite{Safavi-Naeini2013a}, where the integrals are defined as 
$\Bar{I}^{r}=\tilde{\beta}\langle n\rangle^+_\mathrm{eff}$ and 
$\Bar{I}^{b}=\tilde{\beta}(\langle n\rangle^-_\mathrm{eff} + 1)$, with $\tilde{\beta}$ a common prefactor and $\langle n\rangle^\pm_\mathrm{eff}$ the `effective' mode occupations. In the absence of phase noise, these effective mode occupations correspond to the occupation including back-action of the laser.
In the presence of phase noise, they can no longer be interpreted as such, and can be expressed as \cite{Safavi-Naeini2013a}
\begin{align}
    \langle n\rangle^+_\mathrm{eff} &=\frac{\Gamma_m n_{\mathrm{th}}}{\Gamma^{r}_{m,\mathrm{eff}}} - \left( \frac{2\kappa^\mathrm{ext}}{\kappa}\right) \frac{1+C/2}{1+C} n_{\phi}^{\mathrm{photon}}, \\
    \langle n\rangle^-_\mathrm{eff} &=\frac{\Gamma_m n_{\mathrm{th}}}{\Gamma^{b}_{m,\mathrm{eff}}} +\frac{C}{1-C} + \left( \frac{2\kappa^\mathrm{ext}}{\kappa}\right) \frac{1-C/2}{1-C} n_{\phi}^{\mathrm{photon}},
\end{align}
where we assumed again that $\Omega_m \gg \kappa$ and $\kappa=\kappa_1\approx\kappa_2$. If we insert these expressions into $\Bar{I}^r$ and $\Bar{I}^b$ we can find from \cref{eqSIphasenoise:nth_inf} a new relation between the inferred occupancy $n_{\mathrm{th}}^\mathrm{inf}$ and the real occupancy $n_{\mathrm{th}}$ in absence of laser light, which is
\begin{equation}
    \frac{1}{n_{\mathrm{th}}^\mathrm{inf}} = \frac{n_\mathrm{th} +1+\left( \frac{2\kappa^\mathrm{ext}}{\kappa}\right) (1-C/2) n_{\phi}^{\mathrm{photon}} }{n_\mathrm{th}- \left( \frac{2\kappa^\mathrm{ext}}{\kappa}\right) (1+C/2) n_{\phi}^{\mathrm{photon}}} -1 .\label{eqSIphasenoise:nthe-nth}
\end{equation}

We can estimate how laser phase noise affects our thermometry measurements by inverting \cref{eqSIphasenoise:nthe-nth} to express $n_{\mathrm{th}}$ as function of $n_{\mathrm{th}}^\mathrm{inf}$, and assuming typical experimental values ($C=0.15$, $\kappa^{ext} / \kappa \approx 1/2$, $|E_0|^2\approx 6.4\times 10^{14}$ $\mathrm{photons \, s}^{-1}$ and $n_{\mathrm{th}}^\mathrm{inf}=0.4$). This leads to a real occupancy of $n_{\mathrm{th}}=0.407$ or $n_{\mathrm{th}}=0.417$  when using $S^{\mathrm{av}}_\mathrm{\omega\omega}(\Omega_m)$ or $S^{\mathrm{av+std}}_\mathrm{\omega\omega}(\Omega_m)$ for the frequency noise level, respectively.
This shows that the difference between real and inferred occupations is negligible compared to the occupations we measure and their error bars. Similarly small relative differences are found for the occupations of $\sim7$ that we measure at \SI{4}{\kelvin} temperature. 

\subsection{Effective mass}
\label{SI:EffMass}

To calculate the effective mass of the measured phonon modes, we follow the same calculations as Bild \textit{et. al.} \cite{Bild2022} but use updated parameters. In this experiment, the phonon mode we couple to is not strictly an eigenmode of the system, but is instead formed by a superposition of eigenmodes.
The exact superposition is found by maximizing the coupling Hamiltonian
\begin{equation}
    \label{eqSIeffM:OM_interaction_Hamiltonian}
    \hat{H}_\mathrm{int} = \int \dd V \; \epsilon_0 \epsilon_\mathrm{r}^2 \boldsymbol{p} \hat{\boldsymbol{S}}(\Vec{r}) \hat{\Vec{E}}_\mathrm{o,j}(\Vec{r}) \hat{\Vec{E}}_\mathrm{o,j+1}(\Vec{r}),
\end{equation}
where $\boldsymbol{p}$ is the photoelastic tensor, $\hat{\boldsymbol{S}}(\Vec{r})$ is the mechanical strain field, and $\hat{\Vec{E}}_\mathrm{o,j}(\Vec{r})$ and $\hat{\Vec{E}}_\mathrm{o,j+1}(\Vec{r})$ are the two optical modes. Since both electric field modes have approximately identical Gaussian shapes with width $w_0 \approx \SI{77}{\micro\meter}$, the coupling Hamiltonian is an overlap integral of the strain field with an effective field with Gaussian width $w_0/\sqrt{2}$. Thus, the superposition of mechanical eigenmodes will form an effective mode field that has the same width $w_0/\sqrt{2}$. Since the Rayleigh length of the optical modes ($\sim\SI{18.4}{\milli\meter}$ inside Quartz) is larger than the crystal thickness of $L = \SI{5}{\milli\meter}$, we treat the mode field diameter as constant.

Knowing the shape of the mechanical strain field, we can equate the mechanical energy with the potential energy of an effective mechanical mode
\begin{align}
    U &= \frac{c_{33}}{2} \int_V \dd V S_0 \sin(\frac{m \pi z}{L}) e^{-\qty(r/(w_0/\sqrt{2}))^2} \nonumber \\
    &= \frac{1}{2} M_\mathrm{eff} \Omega_m^2 x_\mathrm{eff}^2,
\end{align}
where $c_{33}$ is the relevant stiffness tensor component, $S_0$ is the maximum strain and $m$ the longitudinal mode number. With $\Omega_m = \frac{2\pi c}{\lambda_m}$, $\lambda_m = \frac{2 L}{m}$ and $c = \sqrt{\frac{c_{33}}{\rho}}$, where $\rho$ is the density of the material, this yields for the effective mass
\begin{align}
    M_\mathrm{eff} = \qty(\frac{S_0^2 L^2}{4\pi^2 m^2 x_\mathrm{eff}^2}) \rho \pi \qty(\frac{w_0}{\sqrt{2}})^2 L \label{eqSIeffM:effM}
\end{align}

To calculate the effective mass $M_\mathrm{eff}$, one now has to define what exactly is meant with \emph{effective} displacement $x_\mathrm{eff}$. The displacement in $z$-direction is found by integrating the strain and is given by
\begin{align}
    \label{eqSIeffM:displacement_field}
    u_z (r,\phi,z) = -\frac{L}{m \pi} S_0 \cos(\frac{m \pi z}{L}) e^{-\qty(r/(w_0/\sqrt{2}))^2}
\end{align} We can either choose the effective displacement to be the maximum amplitude of the displacement field $x_{max} = \frac{L}{m \pi} S_0$, or the root mean square (RMS) of the displacement field amplitude. To obtain $x_\mathrm{RMS}$, we have to choose a volume that contains most of the energy of the mode over which to take the RMS, which in our case is a cylinder with volume $\pi R^2 L$, where $R=2 \frac{w_0}{\sqrt{2}}$. So we find
\begin{align}
    x_\mathrm{RMS} &= \sqrt{ \frac{1}{\pi R^2 L} \int_0^L \dd z \int_0^{2\pi} \dd \phi \int_0^R \dd r ~ r \abs{u_z(r,\phi,z)}^2 } \nonumber \\
    &\approx \frac{1}{4} \frac{L}{m \pi} S_0.
\end{align}

This yields the equations for the effective mass
\begin{align}
     M_\mathrm{eff}^{x_{max}} = \frac{1}{4} \rho \pi \qty(\frac{w_0}{\sqrt{2}})^2 L \label{eqSIeffM:effM_xmax} \\
     M_\mathrm{eff}^{x_\mathrm{RMS}} = 4 \rho \pi \qty(\frac{w_0}{\sqrt{2}})^2 L. \label{eqSIeffM:effM_xRMS}
\end{align}
With the density of quartz $\rho = \SI{2.65}{\gram/\centi\meter^3}$, this yields values for the effective mass of the measured phonon mode of $M_\mathrm{eff}^{x_{max}} \approx \SI{31}{\micro\gram}$ and $M_\mathrm{eff}^{x_\mathrm{RMS}} \approx \SI{494}{\micro\gram}$. We argue that for the purposes of this paper, the root mean square displacement is more representative as \textit{effective} displacement.

\end{document}